\documentclass[%
reprint,
superscriptaddress,
 amsmath,amssymb,
 aps,
]{revtex4-2}

\usepackage{graphicx}% Include figure files
\usepackage{dcolumn}% Align table columns on decimal point
\usepackage{bm}% bold math

\usepackage{ulem}
\usepackage{amsmath}

\usepackage{xcolor}

\begin{document}

%=================================================================
% Full title of the paper (Capitalized)
\title{Force Dependence of Proteins’ Transition State Position and the Bell-Evans Model}

% Authors, for the paper (add full first names)
\author{M. Rico-Pasto}
 \thanks{Equally contributed authors}
 \affiliation{%
 Condensed Matter Physics Department, University of Barcelona, C/Marti i Franques 1, 08028 Barcelona, Spain}%

\author{A. Zaltron}
 \thanks{Equally contributed authors}
 \affiliation{%
  Department of Physics and Astronomy, University of Padova, via Marzolo 8, 35131 Padova, Italy}%
 
\author{F. Ritort}
\email[Corresponding author: ]{ritort@ub.edu}
 \affiliation{%
 Condensed Matter Physics Department, University of Barcelona, C/Marti i Franques 1, 08028 Barcelona, Spain}%
 \email{fritort@gmail.com}

% Abstract (Do not insert blank lines, i.e. \\) 
\begin{abstract}
    Single-molecule force spectroscopy has opened a new field of research in molecular biophysics and biochemistry. Pulling experiments on individual proteins permit us to monitor conformational transitions with high temporal resolution and measure their free energy landscape. The force-extension curves of single proteins often present large hysteresis, with unfolding forces that are higher than refolding ones. Therefore, the high energy of the transition state (TS) in these molecules precludes kinetic rates measurements in equilibrium hopping experiments. In irreversible pulling experiments, force dependent kinetic rates measurements show a systematic discrepancy between the sum of the folding and unfolding TS distances derived by the kinetic Bell-Evans model and the full molecular extension predicted by elastic models. Here, we show that this discrepancy originates from the force-induced movement of TS. Specifically, we investigate the highly kinetically stable protein barnase, using pulling experiments and the Bell-Evans model to characterize the position of its kinetic barrier. Experimental results show that while the TS stays at a roughly constant distance relative to the native state, it shifts with force relative to the unfolded state. Interestingly, a conversion of the protein extension into amino acid units shows that the TS position follows the Leffler-Hammond postulate: the higher the force, the lower the number of unzipped amino acids relative to the native state. The results are compared with the quasi-reversible unfolding-folding of a short DNA hairpin.
\end{abstract} 

\maketitle
%%%%%%%%%%%%%%%%%%%%%%%%%%%%%%%%%%%%%%%%%%

\section{Introduction}

A prominent question in biophysics is how biomolecules, and particularly proteins, fold. At present, two debated theories on protein folding are based on the \textit{energy landscape} and \textit{foldon} hypotheses \cite{Frauenfelder1991, Bryngelson1995, Baldwin1995, Maity2005}. The first describes  protein folding as a thermally activated relaxation process in a funneled energy landscape with the native state at the bottom of the funnel \cite{Frauenfelder1991, Bryngelson1995}. The funnel is rugged with deep valleys leading to intermediate states of short lifetime, where the polypeptide chain is partially folded. In the funnel model there are different trajectories for protein folding passing through one or more intermediates. In the late '80s, bulk hydrogen exchange, NMR, and mass spectrometry studies and theoretical models consistently observed recurrent intermediates during folding \cite{Camacho1993, Shaknovich1994, Chan1994, Bai1995}. Based on these results, the foldon hypothesis claims that proteins fold along a unique path in the energy landscape connecting the native and unfolded state through several foldons \cite{Baldwin1995, Maity2005}.

Single-molecule techniques have emerged to investigate the thermodynamics of individual proteins with high temporal resolution \cite{Rief1997, Kellermayer1997}. Force spectroscopy techniques such as Atomic Force Spectroscopy, Magnetic and Optical Tweezers have significantly contributed in achieving new insights into this research area \cite{Neumann2008, Zaltron2020}. Specifically, these methods use force as a "denaturant agent" to mechanically break the bonds that stabilize the protein structure \cite{Bustamante2020}. The force is applied to the N- and C- termini of the polypeptide chain, defining a proper reaction coordinate, the end-to-end distance or molecular extension, useful to describe the folding free energy landscape (mFEL) \cite{Gebhardt2010, Neupane2016, Rebane2016}. Over the years, single-molecule experiments have permitted the reconstruction of the free energy landscape of a wide variety of proteins, both characterized by a two-state folding/unfolding process \cite{Sharma2007, Zheng2014, Goldman2015, Neupane2016} and in the presence of intermediates \cite{Cecconi2005, Gebhardt2010, Stigler2011, Stockmar2016, Yu2017}.

\begin{figure*}
\includegraphics{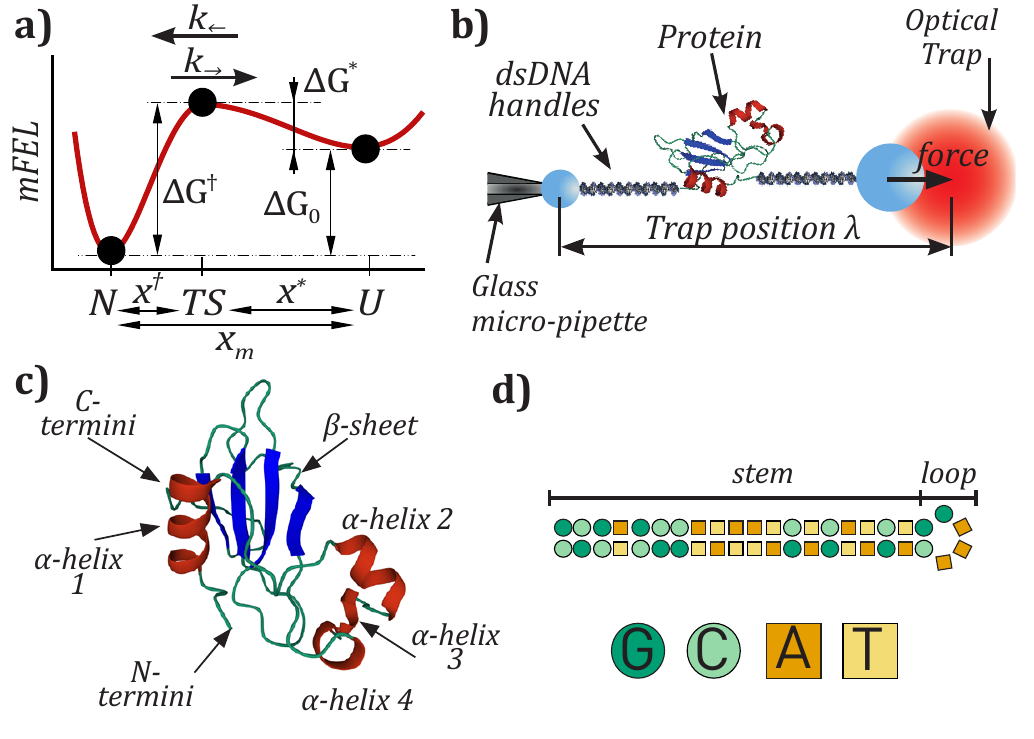} 
\caption{Single-molecule experiments. \textbf{(a)} Illustration of the mFEL of barnase. A kinetic barrier (located at the transition state $TS$) mediates transitions between $N$ and $U$. $\Delta G_0$ is the free energy difference between $N$ and $U$, whereas $\Delta G^\dagger$ and $\Delta G^\ast$ are the kinetic barriers relative to $N$ and $U$, respectively. (\textbf{b}) Illustration of the single-molecule experimental setup. The optical trap measures the force $f$ applied to the ends of the molecular construct. The distance between the optical trap and the pipette, $\lambda$, denotes the control parameter of the experiment. (\textbf{c}) Structure of the protein barnase \cite{Martin1999} (\textbf{d}) Sequence of the DNA hairpin \cite{Forns2011}.}
\label{fig1} 
\end{figure*}

In single-molecule experiments, the mFEL is widely investigated employing the phenomenological Bell-Evans model \cite{Evans1997, Bell1978, Merkel1999, Evans2001}. For a two-state system, the mFEL consists of two wells representing the native ($N$) and the denatured (or unfolded, $U$) states separated by a kinetic barrier placed at the transition state ($TS$) (Figure \ref{fig1}a). The Bell-Evans (BE) model assumes that the position of the barrier relative to $N$ ($x^\dagger$) and $U$ ($x^\ast$) is fixed. In contrast, its height relative to $N$ ($U$) is reduced (increased) upon increasing the applied force in a linear fashion. Moreover, the unfolding and folding kinetic rates, $k_\to$ and $k_\leftarrow$, vary exponentially with the activation energies of the system and, as a result, are force-dependent. The BE model is often used to interpret the data derived either by \textit{equilibrium hopping experiments} and \textit{non-equilibrium pulling experiments}. 

Hopping experiments are used to derive $k_\to$ and $k_\leftarrow$ from equilibrium force-time traces by determining the average lifetimes at each state \cite{Heidarsson2014, Jahn2018, Mehlich2020}. These experiments provide good estimations for the molecular extension, $x_m = x^\dagger + x^\ast$, and the coexistence force $f_c$ (where $k_\to (f_c) = k_{\leftarrow}(f_c)$) for molecules with lifetimes that fall in the experimentally accessible timescales, a requirement often met in DNA molecules but not in proteins. The difference arises from the large  kinetic stability induced by protein tertiary structure. In fact, proteins typically refold at low forces, where molecular conformational transitions are hardly detectable, or are characterized by pronounced hysteresis. Therefore, unfolding and folding events occur in different force ranges, rendering hopping unobservable in experimental times. One might extract kinetic rates from force jump protocols, however these experiments can be inaccurate if the unfolding and folding transitions after the jump occur too fast. In such cases, non-equilibrium pulling experiments are useful to derive the kinetic rates from the survival probabilities of $N$ and $U$, along the unfolding and folding trajectories, respectively \cite{Shank2010, Motlagh2016, Alemany2016, Liu2019}. Indeed, in these experiments, the large molecular folding timescales do not impede measurements of $k_\to$ and $k_\leftarrow$ at different forces. Thus they can be used to investigate also proteins with high kinetic stability. 

It has been shown that the molecular extension $x_m$ derived from the BE model systematically underestimates the predictions based on the elastic properties of proteins \cite{Cecconi2005, Shank2010, Alemany2016}. Thus, the question arises about how to study kinetically stable proteins by combining pulling experiments with the BE model. Therefore, we analyse the unfolding/folding kinetic rates of barnase at room temperature (298\,K), a two-state protein that has been shown to present
pronounced hysteresis upon mechanical folding and unfolding with optical tweezers \cite{Alemany2016}. Barnase is a good candidate to test the BE model's success and limitations to reproduce the folding kinetics in mechanical unzipping experiments. To this aim, we compare the results for barnase with those relative to a reversible folder, such as a DNA hairpin \cite{Forns2011, Pala2011, Rico2018}, which kinetics can be measured in hopping experiments. We find that the sum of $x^\dagger$ and $x^\ast$ derived from pulling experiments underestimates the expected full molecular extension of the protein, $x_m$. However, a conversion of transition state distances into amino acid units permits us to recover the protein extension correctly and, more interestingly, to derive the force dependent folding rates and the coexistence force in the inaccessible intermediate force range. Results are also interpreted in the light of the Leffler-Hammond postulate \cite{Leffler1953, Hammond1955}.

%%%%%%%%%%%%%%%%%%%%%%%%%%%%%%%%%%%%%%%%%%
\section{Materials and Methods}

\subsection{Molecular synthesis}
In single-molecule experiments with optical tweezers, the molecule under investigation is tethered between two beads via double-stranded DNA handles (Figure \ref{fig1}b). The use of DNA handles avoids undesired interactions between the molecule and the beads during the measurements. The molecular construct is connected to two polystyrene beads via specific linkages. The free ends of the DNA handles are labelled with a biotin and a digoxigenin, that bind to beads specifically coated with streptavidin (SA; 2.0–2.9-$\mu$m-diameter bead; G. Kisker Biotech, Steinfurt, Germany) and anti-digoxigenin (AD; 3.0–3.4-$\mu$m-diameter bead; G. Kisker Biotech, Steinfurt, Germany), respectively\cite{Tych2021}. Barnase has been expressed as reported in \cite{Alemany2016}. It contains 110 amino acids (Figure \ref{fig1}c) and folds forming four external $\alpha$-helices surrounding a $\beta$-sheet \cite{Martin1999}); its N- and C-termini have been modified with cysteine-thiol groups that act as anchoring points for the two 500bp dsDNA handles. Finally, the DNA hairpin has been synthesised as described in \cite{Forns2011} and it contains 44 nucleotides that form a 20bp stem ending in a tetraloop (Figure \ref{fig1}d).

\subsection{Optical tweezers setup}
The minitweezers instrument has been described in \cite{Huguet2010}. It consists of two counter-propagating laser beams (845nm) that pass through two microscope objectives (Olympus x60, NA 1.2) forming a single optical trap. The optical trap is moved with nanometric precision using piezoelectric actuators. The experiments are carried out by moving the AD bead captured in the optical trap relative to the SA bead, which is kept fixed at the tip of a glass micro-pipette by air suction (Figure \ref{fig1}b). In this way, an external force is applied to the protein or DNA hairpin tethered between the beads. The instrument records the force and trap position in real-time. Pulling and hopping experiments are carried out at high temporal (1kHz) and nanometer spatial resolution \cite{DeLorenzo2015}.

%%%%%%%%%%%%%%%%%%%%%%%%%%%%%%%%%%%%%%%%%%
\section{Results}
\subsection{Unfolding and folding kinetic rates}\label{kineticrates}
\begin{figure}[ht]
\centering
\includegraphics[width = \linewidth]{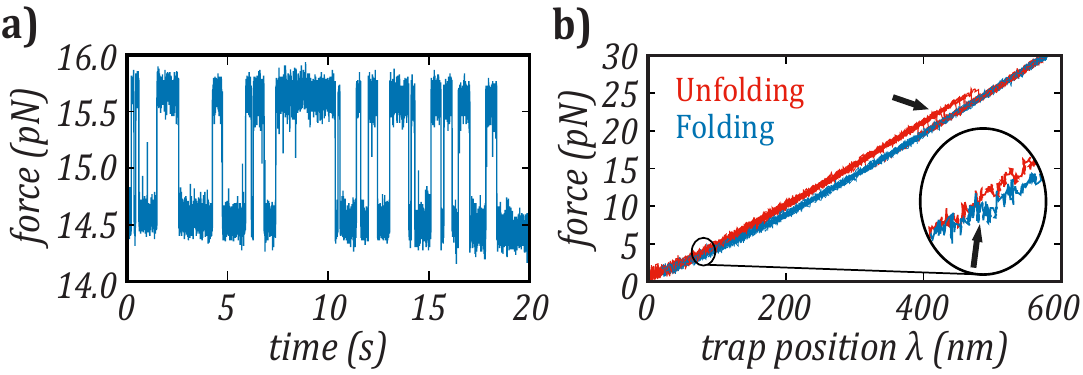}% [width = \linewidth]
\caption{Single molecule experiments on a DNA hairpin and protein barnase. \textbf{(a)} Equilibrium hopping experiments in the 20bp DNA hairpin of Figure\ref{fig1}d: selected trace at the coexistence force ($\sim$ 15pN).  \textbf{(b)} Non-equilibrium pulling experiments on barnase protein. The unfolding (red) and folding (blue) curves present hysteresis, with the folding events occurring at very low forces (below 5\,pN). \label{fig2}}
\end{figure}

We present kinetic rate measurements from pulling experiments on protein barnase, as reported in \cite{Alemany2016}. We compare them with results from DNA hairpins presented in \cite{Forns2011,Rico2018}. The kinetics of DNA hairpins has been measured in equilibrium hopping experiments in the passive mode (see Figure \ref{fig2}a). In these experiments, transitions between different states are monitored while maintaining fixed (clamped) the trap-pipette distance. The kinetic rates are derived from the mean lifetimes of the unfolded and folded states \cite{Tinoco2006}, which have been measured at different trap positions to reconstruct their force dependence. Hopping experiments are widely used in short DNA/RNA hairpins (few tens of base-pairs), since the majority of these molecules fold and unfold in the second and subsecond timescales \cite{Woodside2006, Rico2018} (Figure \ref{fig2}a). Hopping experiments have been also carried out to study conformational transitions in proteins triggered by the reorganization of small domains, e.g., in \cite{Cecconi2005, Gebhardt2010, Gao2011, Gao2012,  Woodside2012}. Still, the folding kinetics of sufficiently large proteins (above a few tens of amino acids) remains innacessible in hopping experiments, making non-equilibrium pulling experiments particularly advantageous. 

\begin{figure}[ht]
\centering
\includegraphics[width = \linewidth]{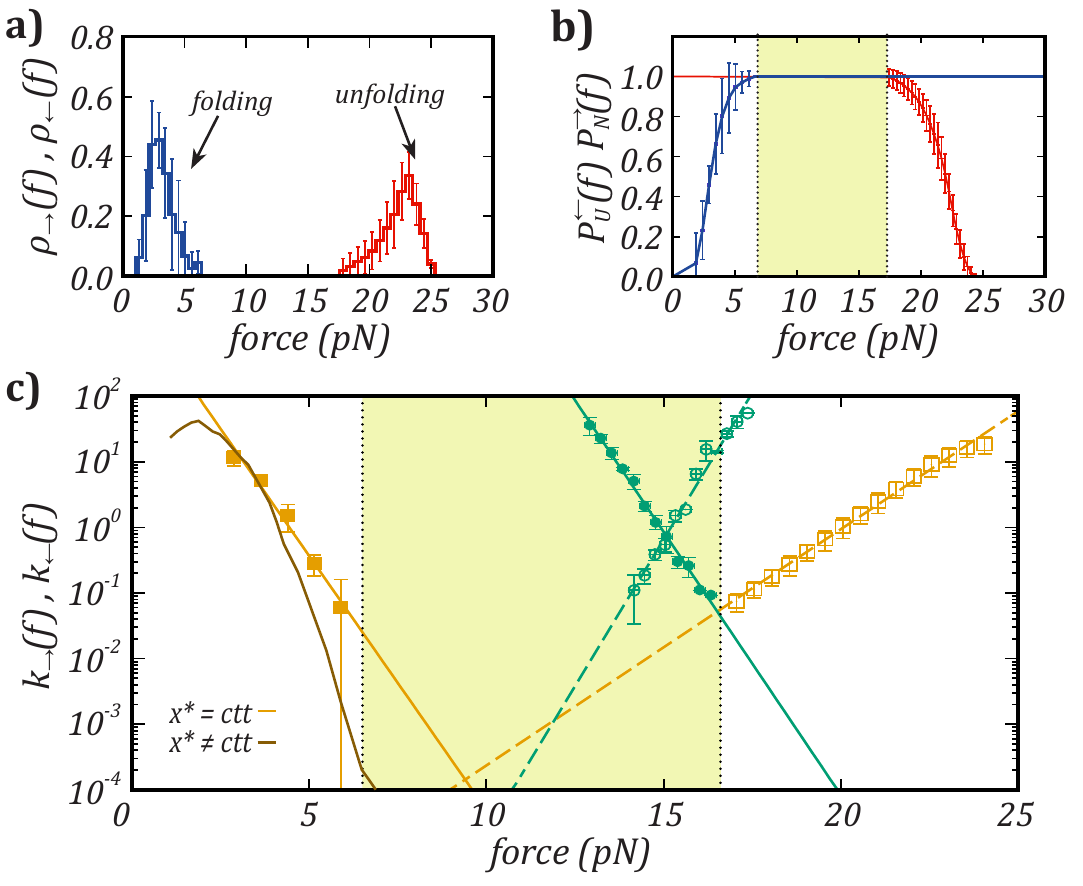}% [width = \linewidth]
\caption{Pulling experiments in barnase at room temperature (298\,K). \textbf{(a)} Barnase unfolding and folding force distributions. \textbf{(b)} Survival probabilities of the folded ($P_U^\leftarrow(f)$, red symbols) and unfolded  ($P_N^\to(f)$, blue symbols) states. \textbf{(c)} Average unfolding (empty symbols) and folding (solid symbols) kinetic rates of the DNA hairpin (green) and barnase (yellow). The yellow area highlights the wide range of forces were the kinetic rates of barnase can not be measured. The yellow and green lines are the fits of the kinetic rates to the BE model; the brown line represents the estimation of the folding kinetic rates of barnase, assuming a movement of the $TS$ position with respect to the unfolded state, $x^\ast$, as discussed in Section \ref{LM} and shown in Figure \ref{fig4}b. \label{fig3}}
\end{figure}

In the pulling experiments carried out in this work, the optical trap is repeatedly moved back and forth at a constant loading rate, $r = 6$pN/s. The force and position are measured, generating the force-distance curves (FDCs) shown in Figure \ref{fig2}b. The protein is pulled from an initial force, $\sim$ 1pN, where it is in $N$, to a final force, $\sim$ 30pN, where it is in $U$, producing unfolding trajectories (red curves in Figure \ref{fig2}b). The unfolding events are observed as sudden force drops in the FDCs (black arrow in Figure \ref{fig2}b, main). When force is relaxed (folding trajectories, blue curve in Figure \ref{fig2}b), the protein folds into $N$ and the event is observed as a sudden force increase of about 0.5\,pN at lower forces, $<5$\,pN (black arrow in the enlarged region of Figure \ref{fig2}b). In \cite{Alemany2016}, it has been shown that the molecular extension upon barnase unfolding corresponds to the release of the 110aa of the native structure, proving that barnase always folds into N in the refolding process. To determine the folding events down to forces as low as 1pN we used a median-filter applied to the unfolding and folding trajectories. The folded branch can be fitted to a linear function, $f(\lambda)= k\lambda$, with $k$ being the effective stiffness of the optical trap plus molecular construct. We used this fit to build a reference folded-baseline by subtracting the folded force branch to the linear fit. The Gaussian distribution of forces along the folded-baseline and a Bayesian method are used to classify into $N$ and $U$ the median-filtered data points along a given folding trajectory. The folding force is defined as the point along the folding trajectory in which the transition $U\to N$ is observed. Even if we can detect folding events below 2pN, we do not detect all of them. For this reason, we have not included in Fig.\ref{fig3}c the force points below 2pN.


From the force-distance curves, we extract the rupture force distributions $\rho_\to(f)$ and $\rho_\leftarrow(f)$ of the first unfolding and folding events, as reported in Figure \ref{fig3}a for a selected molecule. The plots clearly show that the mechanical folding/unfolding of barnase is highly irreversible, with separated ($\sim$ 10\,pN) unfolding and folding force distributions. The large hysteresis impedes to extract kinetic rates $k_\to$ and $k_\leftarrow$ by doing hopping experiments, due to the disparity of lifetimes of states $N$ and $U$. Unfolding and folding kinetic rates of barnase can be derived from pulling experiments by measuring the survival probabilities of states $N$ and $U$, $P_N^\to(f)$ and $P_U^\leftarrow(f)$. These are given by:
\begin{subequations}
\begin{align}
   P^\to_N(f) = 1 - \int_{f_{min}}^f\rho_\to(f') df' \label{eq:PN} \\
   P^\leftarrow_U(f) = 1 - \int_f^{f_{max}}\rho_\leftarrow(f') df' \  \label{eq:PU} 
\end{align}
\end{subequations}
where $f_{min}$ and $f_{max}$ denote the miminum and maximum forces (for simplicity we can take them as $-\infty$ and $\infty$, respectively). Figure \ref{fig3}b shows the survival probabilities $P_N^\to(f)$ and $P_U^\leftarrow(f)$ of protein barnase at 25ºC, where the high irreversibility of its mechanical folding is evidenced. In fact, $P_N^\to(f)$ and $P_U^\leftarrow(f)$ do not cross in the investigated force range (highlighted area in Figure \ref{fig3}b). Finally, by modelling the unfolding and folding kinetics as a first-order Markov process, kinetic rates can be calculated from the rupture force distributions as \cite{Hummer2003},
\begin{subequations}
    \begin{align}
        \rho_\to(f)=  \frac{k_{\rightarrow}(f)}{r}P_{N}^\to(f) \,\, \Rightarrow \,\,
k_{\rightarrow}(f)=  r\,\frac{\rho_{\rightarrow}(f)}{P_{N}(f)} \label{eq:kNU_def}\\
\rho_\leftarrow(f)=  \frac{k_{\leftarrow}(f)}{r}P_{U}^\leftarrow(f)\,\, \Rightarrow \,\,
k_{\leftarrow}(f)= r\,\frac{\rho_{\leftarrow}(f)}{P_{U}^\leftarrow(f)} \label{eq:kUN_def}
    \end{align}
\end{subequations}
where $r = |df/dt|$ is the constant loading and unloading rate. The results are shown in a log-normal plot in Figure \ref{fig3}c (yellow symbols), where kinetic rates for the DNA hairpin have been also reported for comparison (green symbols). In this case, folding and unfolding kinetic rates span a narrow range of forces ($\sim$ 13\,-\,17\,pN), around the coexistence force ($\sim$ 15\,pN). In contrast, the yellow area highlights the force range where kinetic measurements from equilibrium hopping experiments are not feasible for barnase. Kinetic rates of the DNA hairpin and barnase present a linear force-dependence, which is discussed in the next section.

\subsection{The transition state position and the Leffler-Hammond postulate}\label{LM}

The force dependence of the folding and unfolding kinetic rates is well approximated by the BE model. In this model, the height of the barrier located at the $TS$ linearly decreases with the applied force $f$. The decrease rate equals the distances of the $TS$ to the native and denatured states ($x^\dagger$ and $x^\ast$) which are taken as constant. The BE kinetic rates are given by:

\begin{subequations}
\begin{align}
   k_\to (f) = k_a \exp{\big(-\beta \Delta G^\dagger\big)} \cdot \exp{\Big( \beta f x^\dagger \Big)}\label{eq:kNU} \\
    k_\leftarrow (f) = k_a \exp{\big( -\beta \Delta G^\ast \big)} \cdot \exp{ \Big( \beta \big( \Delta G_0 - f x^\ast \big) \Big)} \ \label{eq:kUN} 
\end{align}
\end{subequations}
where $\Delta G_0$ is the free energy difference between $N$ and $U$, $\Delta G^\dagger$ and $\Delta G^\ast$ are the barrier's heights relative to $N$ and $U$ (see Figure \ref{fig1}a), $\beta = 1/k_BT$ (with $k_B$ the Boltzmann constant and $T$ the temperature), and $k_a$ is the attempt rate, which depends on the experimental conditions.

By fitting Eqs. \eqref{eq:kNU} and \eqref{eq:kUN} to the experimental values of $k_\to(f)$ and $k_\leftarrow(f)$ (dashed and solid lines in Figure \ref{fig3}c), we derive the position and the height of the energetic barriers. First, the values that best fit the BE model to the experimental data have been used to estimate the coexistence force for barnase $f_c = (9.0\pm 0.5)$\,pN, where $k_\to(f_c)\,=\,k_\leftarrow(f_c)$. The error in the estimate includes the systematic error in force calibration which is about 5\%. $N$ and $U$ have the same free energy at the coexistence force, meaning that the barriers to unfold or fold the protein are the same. Second, we have carried out an in-depth analysis about the position of the kinetic barrier. From the fits to the kinetic rates, we have derived $x^\dagger$ and $x^\ast$. The errors of $x^\dagger$ and $x^*$ have been calculated as the standard statistical error taken from the values taken over all five studied molecules, giving $x^\dagger = (3\pm 1)$nm and $x^\ast= (8\pm 1)$nm, which sum $x_{m,exp} = x^\dagger + x^\ast = (11 \pm 2)$nm. In the BE model, the latter should be equal to the full molecular extension of the protein at the coexistence force. In the following, we show that $x_{m,exp} =  (11 \pm 2)$nm largely underestimates the value predicted from the elastic theory of polymers. This indicates that the assumption that $x^\dagger,x^\ast$ are force-independent does not hold. Moreover, in \cite{Alemany2016}, the unfolding and folding kinetic rates have been studied with the Dudko-Hummer-Szabo (DHS) model \cite{Dudko2006}. This model is parametrized by a parameter $\gamma$ that interpolates between different TS shapes (BE corresponding to $\gamma=1$) finding $x_{m,exp} =(10\pm2)$nm for a parabolic shape ($\gamma=1/2$) and $x_{m,exp} =(10\pm1)$nm for a cubic shape ($\gamma=2/3$), whereas for the BE model we get $x_{m,exp} =(11\pm2)$nm. Considering the expected value from the WLC prediction, $x_{m,th} =(21\pm1)$nm, the above results show that both BE and DHS largely underestimate the full molecular extension.

To derive the theoretical prediction for $x_{m,th}(f)$, we have calculated the extension of the polypeptide chain released upon unfolding at a given force $f$, $X(f)$. To $X(f)$ we must subtract the protein elongation in the native state just before the rip, $x_d(f)$, which gives $x_{m,th}(f) = X(f) - x_d(f)$. The initial extension of the folded protein, $x_d(f)$, is modeled as a dipole of length equal to the N-C-termini distance (3nm for barnase \cite{Martin1999, Alemany2016}). The dipole orients under an applied force, its average extension being equivalent to that of a single Kühn segment (in the Freely Jointed Chain (FJC) model) of length equal to that of the dipole (3nm). Besides, the value of $X(f)$ is determined by using the inextensible Worm-Like-Chain (WLC) model and its interpolation formula between high and low forces \cite{Siggia1994, ViaderGodoy2021}
\begin{equation}\label{eq:iWLC}
    f = \frac{k_BT}{4 L_p} \left[  \left(1-\frac{X(f)}{N d_{aa}}\right)^{-2} + 4\frac{X(f)}{N d_{aa}} -1 \right]
\end{equation}
where $L_p=0.8$nm is the persistence length of the polypeptide chain, $N$ is the number of residues (110 for barnase), and $d_{aa}=0.37$nm is the aa-distance \cite{Alemany2016}. Using these elastic parameters, the theoretical molecular extension at $f_c$ is $x_{m,th}(f_c)= X(f_c) - x_d(f_c) = (21 \pm 1)$nm, which differs by $\sim$ 9nm from the value derived from the BE model, $x_{m,exp} =  (11 \pm 2)$nm. This discrepancy arises from the strong hysteresis between unfolding and folding and the different range of forces used to measure $x^\dagger$ and $x^\ast$. Indeed, fits of $k_\to$ and $k_\leftarrow$ to the BE model are made close to the most probable unfolding and refolding forces, meaning that the values of $x^\dagger$ and $x^\ast$ are estimated at very different forces, far from $f_c$. In contrast, the expected molecular extension has been calculated from the elastic models at $f_c$. Therefore, to properly compare the measured values $x^\dagger$ and $x^\ast$ with the theoretical prediction, these must be estimated at the same force. In the DNA hairpin, both unfolding and folding kinetic rates have been measured in the same force range close to $f_{c, DNA} \sim 15$\,pN. As a consequence, the DNA extension theoretically predicted by the WLC model, $x_{DNA,th}  = ( 18.0 \pm 0.3 )$nm (with $L_p$, $d_{aa}$ and $x_d$ given in \cite{alemany2014, ViaderGodoy2021}) matches that obtained from the BE model ($x_{DNA,exp}  = ( 17.9 \pm 0.4 )$nm). The number of released bases at the transition state at $f_c$ are $n^\dagger = 20\pm1$ and $n^\ast = 23\pm2$, which sum gives the total $44$ bases of the hairpin.

To estimate the $TS$ position of barnase, we have converted barrier distances $x^\dagger$ and $x^\ast$ into the number of unfolded amino acids relative to $N$ ($n^\dagger_{aa}$) and $U$ ($n^\ast_{aa}$), by using the WLC model, Eq. \eqref{eq:iWLC}. The obtained values at the most probable unfolding and folding forces are $n^\dagger_{aa}= (13\pm1)$aa and $n^\ast_{aa} = (69 \pm 3)$aa, which sum $n^\dagger_{aa}+n^\ast_{aa}\simeq 82$ underestimates the full number of amino acids (110).
\begin{figure}[th]
\centering
\includegraphics[width = \linewidth]{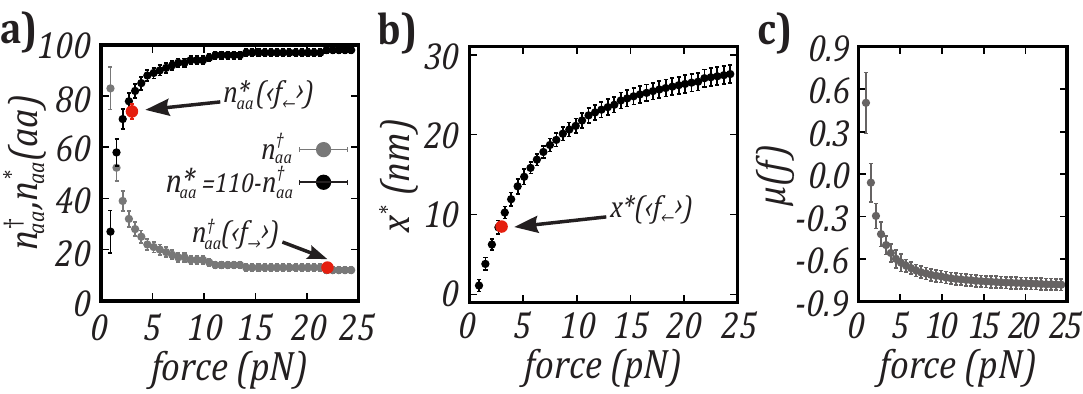}%[width = \linewidth]
\caption{Force-dependent \textit{TS} position. \textbf{(a)} Number of released and absorbed amino acids at $TS$ relative to $N$, $n_{aa}^\dagger$ (grey), and $U$, $n_{aa}^\ast$ (black), respectively. The force dependence of $n_{aa}^\dagger$ is evaluated assuming a constant $x^\dagger$ and using the WLC model; the values of $n_{aa}^\ast$ are calculated as $110 - n_{aa}^\dagger$. The red circles show the values of $n_{aa}^\dagger$ and $n_{aa}^\ast$ calculated at the most probable unfolding and refolding forces, by fitting the BE model to the experimental kinetic rates. \textbf{(b)} $TS$ position relative to $U$, $x^\ast$, versus force. The values of $x^\ast$ are derived by using Eq. \eqref{eq:iWLC} and the values of $n_{aa}^\ast$ reported in (a). The red circle represents the value of $x^\ast$ calculated at the most probable folding force, by using the experimental folding kinetic rates. \textbf{(c)} Fragility of barnase versus force. \label{fig4}}
\end{figure} 
This result could indicate the presence of intermediate states in the unfolding/folding pathway. However, calorimetric studies \cite{Makarov1993, Griko1994} and numerical simulations \cite{Galano2019} have previously demonstrated that barnase folds in a two-state manner. A plausible explanation of the discrepancy is that the distances to the $TS$ (expressed in amino acid units) are not fixed but change with force. In fact, upon increasing the force, the extension of that part of the polypeptide chain that is unfolded at $TS$ (which contains $n_{aa}^\dagger$ aa) should increase as predicted by Eq. \eqref{eq:iWLC}. At the same time, $n_{aa}^\dagger$ decreases with force as predicted by the Leffler-Hammond postulate for chemical reactions \cite{Leffler1953,  Hammond1955}. According to this postulate, upon applying an external perturbation (e.g., force), the $TS$ position moves toward $N$ ($U$) to counteract the increased thermodynamic stability of the $U$ ($N$) state. Upon stretching, the postulate predicts that the number of released (adsorbed) amino acids in the $TS$ during unfolding (folding), $n_{aa}^\dagger$ ($n_{aa}^\ast$), decreases (increases) with force. In the unfolding case, a compensation between the two effects leads to a $N-TS$ extension $x^\dagger$  which is roughly constant, explaining the linearity of the $\log(k_{\rightarrow})$ versus force observed in Figure \ref{fig3}c in the range 16-23pN (yellow dashed line). In contrast, such linearity is not expected for the folding rate ($\log(k_{\rightarrow})$ versus force). Upon increasing the force, $n_{aa}^\ast$ increases with force, therefore no compensation occurs between such increase and that of the molecular extension, $x^\ast$. 

To verify this, we have used the WLC model (Eq. \eqref{eq:iWLC}) to calculate $n_{aa}^\dagger$ over the full range of forces, assuming a constant value of $x^\dagger$. The results are reported in Figure \ref{fig4}a and show that $n_{aa}^\dagger$ decreases when the force increases, in agreement with the Leffler-Hammond postulate. Then, we used the value of $n_{aa}^\dagger$ estimated at the most probable folding force $\langle f_\leftarrow \rangle$ = ($3.0 \pm 0.2$)\,pN to sum it with the value of $n_{aa}^\ast$ obtained from the folding kinetic rates and the WLC model. At $\langle f_\leftarrow \rangle$, the protein extension in amino acids is $n_{aa}^{tot} = n_{aa}^\dagger + n_{aa}^\ast = (37\pm 1) + (69 \pm 3)$aa = $106 \pm 4$aa, a result compatible with the 110 amino acids of barnase. This result confirms the two-state behavior of barnase folding and, above all, the effective movement of the transition state which occurs to counteract the action of the force. The compensation between the change in $n_{aa}^\dagger$ and the molecular extension with force in \eqref{eq:iWLC}, reported at forces as low as $\langle f_\leftarrow \rangle$ = ($3.0 \pm 0.2$)\,pN shows that the linear behavior of log($k_{\rightarrow}$) versus $f$ leads to an approximately constant value of $x^\dagger$ throughout the whole range of forces. Probably this compensation is just casual, in fact a constant $x^\dagger$ is not always observed, e.g., in DNA and RNA hairpins \cite{Changbong2006, Manosas2006, Alemany2017}. 

From $n_{aa}^\dagger(f)$ we extract $n_{aa}^\ast(f)$ by subtracting $n_{aa}^\dagger(f)$ to the total number of amino acids (110), $n_{aa}^\ast(f)=110-n_{aa}^\dagger(f)$. Figure \ref{fig4}a shows that $n_{aa}^\ast(f)$ exhibits a force dependence opposite to that of $n_{aa}^\dagger$, approaching $U$ as force decreases. In Figure \ref{fig4}b, we show $x^\ast(f)$, derived by using Eq. \eqref{eq:iWLC} and the values of $n^\ast_{aa}$ reported in Figure \ref{fig4}a. In contrast to $x^\dagger$, $x^\ast$ presents a strong force dependence. Its value increases up to $\sim 30$\,nm at the maximum unfolding force ($\sim$25\,pN). Another interesting quantity is the mechanical fragility of barnase, defined as $\mu(f)=(x^\dagger-x^\ast)/x_m$, where $x_m=x^\dagger+x^\ast$. In Figure \ref{fig4}c we show the fragility as a function of force. The Leffler-Hammond postulate predicts that the fragility is a monotonically decreasing function of force \cite{Alemany2017}, with compliant unfolding at low forces and brittle unfolding at high forces. 

%%%%%%%%%%%%%%%%%%%%%%%%%%%%%%%%%%%%%%%%%%%%%%%%%%%%%%%%%
\section{Discussion}

In this work, we have investigated the effect of force on the molecular free energy landscape (mFEL) of protein barnase. Barnase is a suitable model due to its high mechanical stability showing large hysteresis in the FDCs \cite{Alemany2016}. Barnase also shows a two-state folding, in bulk experiments\cite{Makarov1993, Griko1994} and simulations\cite{Galano2019}. In particular, we address the question of the reported underestimation of the molecular extension  when the BE model is used to fit the force-dependent kinetic rates \cite{Gebhardt2010, Motlagh2016, Jahn2018}. 

To this end, we have analyzed non-equilibrium pulling data to extract the force-dependent unfolding and folding kinetic rates ($k_\to$ and $k_\leftarrow$) at room temperature. The distances of the transition state ($TS$) to the native ($N$) and unfolded ($U$) states, $x^\dagger$ and  $x^\ast$, have been calculated from the force-dependent kinetic rates using the BE model. The experimentally derived sum $x^\dagger + x^\ast$ underestimates the total molecular extension $x_m$ calculated from the elastic worm-like chain (WLC) model. To resolve this discrepancy, we derived the number of released amino acids at the $TS$, $n_{aa}^\dagger$, from the extensions $x^\dagger$ using the WLC model. We have found that the $TS$ position approaches $N$ upon increasing the force, while $x^\dagger$ remains approximately constant as assumed in the BE model. This result agrees with the Leffler-Hammond postulate, which states that in chemical reactions the $TS$ moves toward the reactants when an external agent favors the products \cite{Leffler1953, Hammond1955}. Nuclear Magnetic Resonance (NMR) spectroscopy experiments in the early '90s confirmed the validity of this postulate using engineered mutants to modify the thermodynamic stability of barnase \cite{Matouschek1993}. Our results confirm that force is a very efficient denaturant, capable of inducing a movement of the $TS$. The results of Figure\ref{fig4}a confirm that $x^\dagger(f)$ is roughly constant due to the compensation between the decrease of $n_{aa}^\dagger(f)$ and the increase of the molecular extension per aa, ($x^\dagger_{1aa}(f)$) with force $f$, or $x^\dagger(f)=n_{aa}^\dagger(f)x^\dagger_{1aa}(f)\simeq {\rm cnst}$. In contrast, such compensation does not occur for $x^\ast(f)$ which markedly changes with force (Figure \ref{fig4}b), predicting a curvature in the $\log (k_\leftarrow)$ versus $f$ plots (Figure \ref{fig3}c, brown line). Yet, the curvature observed in the data (filled squares in Figure \ref{fig3}c), is smaller than that predicted from $x^\ast(f)$ in Figure \ref{fig4}b. This discrepancy indicates that the assumption of a constant $x^\dagger(f)$ is only an approximation that might be refined by matching the measured and predicted values of $k_\leftarrow(f)$. 

Finally, one might ask about the biological significance of the variation of $n_{aa}^\dagger$ with force reported in Figure \ref{fig4}, which shows a marked decrease below 5\,pN. This result might indicate a regulatory role of force in a wide variety of biological processes governed by conformational changes of proteins. Our results confirm the general validity of the BE model to investigate the kinetics and mFEL of highly kinetically stable proteins. The analysis based on the movement of the $TS$ position paves the way for future studies on other proteins with a mechanical role, such as scleroproteins \cite{Waite1983}, metalloproteins that bind to metal ions \cite{Lu2009} or lipoproteins, which are transient intermediates in the process of lipids' transfer \cite{Hayashi1990}.

%%%%%%%%%%%%%%%%%%%%%%%%%%%%%%%%%%%%%%%%%%
\section*{author contributions} 
Conceptualization, Marc Rico-Pasto, Annamaria Zaltron and Felix Ritort; Data curation, Marc Rico-Pasto; Funding acquisition, Annamaria Zaltron and Felix Ritort; Investigation, Marc Rico-Pasto and Annamaria Zaltron; Supervision, Felix Ritort; Visualization, Marc Rico-Pasto; Writing – original draft, Marc Rico-Pasto and Annamaria Zaltron; Writing – review \& editing, Annamaria Zaltron and Felix Ritort.

\section*{funding} 
M.R-P. and F.R. authors  acknowledge financial support from Grants Proseqo (FP7 EU program) FIS2016-80458-P (Spanish Research Council), Icrea Academia prizes 2013 and 2018 (Catalan Government) and the Spanish Research Council Grant No PID2019-111148GB-I00. A.Z. acknowledges funding from the University of Padua (Supply Award 2015 nº 6710927028 and the grant BIRD207923) and Fondazione Cariparo (grant Visiting Programme 2018 - TIRES)


\bibliography{bibliography}

\begin{thebibliography}{63}%
\makeatletter
\providecommand \@ifxundefined [1]{%
 \@ifx{#1\undefined}
}%
\providecommand \@ifnum [1]{%
 \ifnum #1\expandafter \@firstoftwo
 \else \expandafter \@secondoftwo
 \fi
}%
\providecommand \@ifx [1]{%
 \ifx #1\expandafter \@firstoftwo
 \else \expandafter \@secondoftwo
 \fi
}%
\providecommand \natexlab [1]{#1}%
\providecommand \enquote  [1]{``#1''}%
\providecommand \bibnamefont  [1]{#1}%
\providecommand \bibfnamefont [1]{#1}%
\providecommand \citenamefont [1]{#1}%
\providecommand \href@noop [0]{\@secondoftwo}%
\providecommand \href [0]{\begingroup \@sanitize@url \@href}%
\providecommand \@href[1]{\@@startlink{#1}\@@href}%
\providecommand \@@href[1]{\endgroup#1\@@endlink}%
\providecommand \@sanitize@url [0]{\catcode `\\12\catcode `\$12\catcode
  `\&12\catcode `\#12\catcode `\^12\catcode `\_12\catcode `\%12\relax}%
\providecommand \@@startlink[1]{}%
\providecommand \@@endlink[0]{}%
\providecommand \url  [0]{\begingroup\@sanitize@url \@url }%
\providecommand \@url [1]{\endgroup\@href {#1}{\urlprefix }}%
\providecommand \urlprefix  [0]{URL }%
\providecommand \Eprint [0]{\href }%
\providecommand \doibase [0]{https://doi.org/}%
\providecommand \selectlanguage [0]{\@gobble}%
\providecommand \bibinfo  [0]{\@secondoftwo}%
\providecommand \bibfield  [0]{\@secondoftwo}%
\providecommand \translation [1]{[#1]}%
\providecommand \BibitemOpen [0]{}%
\providecommand \bibitemStop [0]{}%
\providecommand \bibitemNoStop [0]{.\EOS\space}%
\providecommand \EOS [0]{\spacefactor3000\relax}%
\providecommand \BibitemShut  [1]{\csname bibitem#1\endcsname}%
\let\auto@bib@innerbib\@empty
%</preamble>
\bibitem [{\citenamefont {Frauenfelder}\ \emph {et~al.}(1991)\citenamefont
  {Frauenfelder}, \citenamefont {Sligar},\ and\ \citenamefont
  {Wolynes}}]{Frauenfelder1991}%
  \BibitemOpen
  \bibfield  {author} {\bibinfo {author} {\bibfnamefont {H.}~\bibnamefont
  {Frauenfelder}}, \bibinfo {author} {\bibfnamefont {S.}~\bibnamefont
  {Sligar}},\ and\ \bibinfo {author} {\bibfnamefont {P.}~\bibnamefont
  {Wolynes}},\ }\bibfield  {title} {\bibinfo {title} {The energy landscapes and
  motions of proteins},\ }\href {https://doi.org/10.1126/science.1749933}
  {\bibfield  {journal} {\bibinfo  {journal} {Science}\ }\textbf {\bibinfo
  {volume} {254}},\ \bibinfo {pages} {1598} (\bibinfo {year}
  {1991})}\BibitemShut {NoStop}%
\bibitem [{\citenamefont {Bryngelson}\ \emph {et~al.}(1995)\citenamefont
  {Bryngelson}, \citenamefont {Onuchic}, \citenamefont {Socci},\ and\
  \citenamefont {Wolynes}}]{Bryngelson1995}%
  \BibitemOpen
  \bibfield  {author} {\bibinfo {author} {\bibfnamefont {J.~D.}\ \bibnamefont
  {Bryngelson}}, \bibinfo {author} {\bibfnamefont {J.~N.}\ \bibnamefont
  {Onuchic}}, \bibinfo {author} {\bibfnamefont {N.~D.}\ \bibnamefont {Socci}},\
  and\ \bibinfo {author} {\bibfnamefont {P.~G.}\ \bibnamefont {Wolynes}},\
  }\bibfield  {title} {\bibinfo {title} {Funnels, pathways, and the energy
  landscape of protein folding: A synthesis},\ }\href
  {https://doi.org/https://doi.org/10.1002/prot.340210302} {\bibfield
  {journal} {\bibinfo  {journal} {Proteins: Structure, Function, and
  Bioinformatics}\ }\textbf {\bibinfo {volume} {21}},\ \bibinfo {pages} {167}
  (\bibinfo {year} {1995})}\BibitemShut {NoStop}%
\bibitem [{\citenamefont {Baldwin}(1995)}]{Baldwin1995}%
  \BibitemOpen
  \bibfield  {author} {\bibinfo {author} {\bibfnamefont {R.~L.}\ \bibnamefont
  {Baldwin}},\ }\bibfield  {title} {\bibinfo {title} {The nature of protein
  folding pathways: The classical versus the new view},\ }\href
  {https://doi.org/10.1007/BF00208801} {\bibfield  {journal} {\bibinfo
  {journal} {Journal of Biomolecular NMR}\ }\textbf {\bibinfo {volume} {5}},\
  \bibinfo {pages} {103 } (\bibinfo {year} {1995})}\BibitemShut {NoStop}%
\bibitem [{\citenamefont {Maity}\ \emph {et~al.}(2005)\citenamefont {Maity},
  \citenamefont {Maity}, \citenamefont {Krishna}, \citenamefont {Mayne},\ and\
  \citenamefont {Englander}}]{Maity2005}%
  \BibitemOpen
  \bibfield  {author} {\bibinfo {author} {\bibfnamefont {H.}~\bibnamefont
  {Maity}}, \bibinfo {author} {\bibfnamefont {M.}~\bibnamefont {Maity}},
  \bibinfo {author} {\bibfnamefont {M.~M.~G.}\ \bibnamefont {Krishna}},
  \bibinfo {author} {\bibfnamefont {L.}~\bibnamefont {Mayne}},\ and\ \bibinfo
  {author} {\bibfnamefont {S.~W.}\ \bibnamefont {Englander}},\ }\bibfield
  {title} {\bibinfo {title} {Protein folding: The stepwise assembly of foldon
  units},\ }\href {https://doi.org/10.1073/pnas.0501043102} {\bibfield
  {journal} {\bibinfo  {journal} {Proceedings of the National Academy of
  Sciences}\ }\textbf {\bibinfo {volume} {102}},\ \bibinfo {pages} {4741}
  (\bibinfo {year} {2005})}\BibitemShut {NoStop}%
\bibitem [{\citenamefont {Camacho}\ and\ \citenamefont
  {Thirumalai}(1993)}]{Camacho1993}%
  \BibitemOpen
  \bibfield  {author} {\bibinfo {author} {\bibfnamefont {C.~J.}\ \bibnamefont
  {Camacho}}\ and\ \bibinfo {author} {\bibfnamefont {D.}~\bibnamefont
  {Thirumalai}},\ }\bibfield  {title} {\bibinfo {title} {Kinetics and
  thermodynamics of folding in model proteins},\ }\href
  {https://doi.org/10.1073/pnas.90.13.6369} {\bibfield  {journal} {\bibinfo
  {journal} {Proceedings of the National Academy of Sciences}\ }\textbf
  {\bibinfo {volume} {90}},\ \bibinfo {pages} {6369} (\bibinfo {year}
  {1993})}\BibitemShut {NoStop}%
\bibitem [{\citenamefont {Shakhnovich}(1994)}]{Shaknovich1994}%
  \BibitemOpen
  \bibfield  {author} {\bibinfo {author} {\bibfnamefont {E.~I.}\ \bibnamefont
  {Shakhnovich}},\ }\bibfield  {title} {\bibinfo {title} {Proteins with
  selected sequences fold into unique native conformation},\ }\href
  {https://doi.org/10.1103/PhysRevLett.72.3907} {\bibfield  {journal} {\bibinfo
   {journal} {Physical Review Letters}\ }\textbf {\bibinfo {volume} {72}},\
  \bibinfo {pages} {3907} (\bibinfo {year} {1994})}\BibitemShut {NoStop}%
\bibitem [{\citenamefont {Chan}\ and\ \citenamefont {Dill}(1994)}]{Chan1994}%
  \BibitemOpen
  \bibfield  {author} {\bibinfo {author} {\bibfnamefont {H.~S.}\ \bibnamefont
  {Chan}}\ and\ \bibinfo {author} {\bibfnamefont {K.~A.}\ \bibnamefont
  {Dill}},\ }\bibfield  {title} {\bibinfo {title} {Transition states and
  folding dynamics of proteins and heteropolymers},\ }\href
  {https://doi.org/10.1063/1.466677} {\bibfield  {journal} {\bibinfo  {journal}
  {The Journal of Chemical Physics}\ }\textbf {\bibinfo {volume} {100}},\
  \bibinfo {pages} {9238} (\bibinfo {year} {1994})}\BibitemShut {NoStop}%
\bibitem [{\citenamefont {Bai}\ \emph {et~al.}(1995)\citenamefont {Bai},
  \citenamefont {Sosnick}, \citenamefont {Mayne},\ and\ \citenamefont
  {Englander}}]{Bai1995}%
  \BibitemOpen
  \bibfield  {author} {\bibinfo {author} {\bibfnamefont {Y.}~\bibnamefont
  {Bai}}, \bibinfo {author} {\bibfnamefont {T.}~\bibnamefont {Sosnick}},
  \bibinfo {author} {\bibfnamefont {L.}~\bibnamefont {Mayne}},\ and\ \bibinfo
  {author} {\bibfnamefont {S.}~\bibnamefont {Englander}},\ }\bibfield  {title}
  {\bibinfo {title} {Protein folding intermediates: native-state hydrogen
  exchange},\ }\href {https://doi.org/10.1126/science.7618079} {\bibfield
  {journal} {\bibinfo  {journal} {Science}\ }\textbf {\bibinfo {volume}
  {269}},\ \bibinfo {pages} {192} (\bibinfo {year} {1995})}\BibitemShut
  {NoStop}%
\bibitem [{\citenamefont {Rief}\ \emph {et~al.}(1997)\citenamefont {Rief},
  \citenamefont {Gautel}, \citenamefont {Oesterhelt}, \citenamefont
  {Fernandez},\ and\ \citenamefont {Gaub}}]{Rief1997}%
  \BibitemOpen
  \bibfield  {author} {\bibinfo {author} {\bibfnamefont {M.}~\bibnamefont
  {Rief}}, \bibinfo {author} {\bibfnamefont {M.}~\bibnamefont {Gautel}},
  \bibinfo {author} {\bibfnamefont {F.}~\bibnamefont {Oesterhelt}}, \bibinfo
  {author} {\bibfnamefont {J.~M.}\ \bibnamefont {Fernandez}},\ and\ \bibinfo
  {author} {\bibfnamefont {H.~E.}\ \bibnamefont {Gaub}},\ }\bibfield  {title}
  {\bibinfo {title} {Reversible unfolding of individual titin immunoglobulin
  domains by afm},\ }\href {https://doi.org/10.1126/science.276.5315.1109}
  {\bibfield  {journal} {\bibinfo  {journal} {Science}\ }\textbf {\bibinfo
  {volume} {276}},\ \bibinfo {pages} {1109} (\bibinfo {year}
  {1997})}\BibitemShut {NoStop}%
\bibitem [{\citenamefont {Kellermayer}\ \emph {et~al.}(1997)\citenamefont
  {Kellermayer}, \citenamefont {Smith}, \citenamefont {Granzier},\ and\
  \citenamefont {Bustamante}}]{Kellermayer1997}%
  \BibitemOpen
  \bibfield  {author} {\bibinfo {author} {\bibfnamefont {M.~S.~Z.}\
  \bibnamefont {Kellermayer}}, \bibinfo {author} {\bibfnamefont {S.~B.}\
  \bibnamefont {Smith}}, \bibinfo {author} {\bibfnamefont {H.~L.}\ \bibnamefont
  {Granzier}},\ and\ \bibinfo {author} {\bibfnamefont {C.}~\bibnamefont
  {Bustamante}},\ }\bibfield  {title} {\bibinfo {title} {Folding-unfolding
  transitions in single titin molecules characterized with laser tweezers},\
  }\href {https://doi.org/10.1126/science.276.5315.1112} {\bibfield  {journal}
  {\bibinfo  {journal} {Science}\ }\textbf {\bibinfo {volume} {276}},\ \bibinfo
  {pages} {1112} (\bibinfo {year} {1997})}\BibitemShut {NoStop}%
\bibitem [{\citenamefont {Neumann}\ and\ \citenamefont
  {Nagy}(2008)}]{Neumann2008}%
  \BibitemOpen
  \bibfield  {author} {\bibinfo {author} {\bibfnamefont {C.~K.}\ \bibnamefont
  {Neumann}}\ and\ \bibinfo {author} {\bibfnamefont {A.}~\bibnamefont {Nagy}},\
  }\bibfield  {title} {\bibinfo {title} {Single-molecule force spectroscopy:
  optical tweezers, magnetic tweezers and atomic force microscopy},\
  }\href@noop {} {\bibfield  {journal} {\bibinfo  {journal} {Nature Methods}\
  }\textbf {\bibinfo {volume} {5}},\ \bibinfo {pages} {491} (\bibinfo {year}
  {2008})}\BibitemShut {NoStop}%
\bibitem [{\citenamefont {Zaltron}\ \emph {et~al.}(2020)\citenamefont
  {Zaltron}, \citenamefont {Merano}, \citenamefont {Mistura}, \citenamefont
  {Sada},\ and\ \citenamefont {Seno}}]{Zaltron2020}%
  \BibitemOpen
  \bibfield  {author} {\bibinfo {author} {\bibfnamefont {A.}~\bibnamefont
  {Zaltron}}, \bibinfo {author} {\bibfnamefont {M.}~\bibnamefont {Merano}},
  \bibinfo {author} {\bibfnamefont {G.}~\bibnamefont {Mistura}}, \bibinfo
  {author} {\bibfnamefont {C.}~\bibnamefont {Sada}},\ and\ \bibinfo {author}
  {\bibfnamefont {F.}~\bibnamefont {Seno}},\ }\bibfield  {title} {\bibinfo
  {title} {Optical tweezers in single-molecule experiments},\ }\href@noop {}
  {\bibfield  {journal} {\bibinfo  {journal} {European Physical journal Plus}\
  }\textbf {\bibinfo {volume} {135}},\ \bibinfo {pages} {1} (\bibinfo {year}
  {2020})}\BibitemShut {NoStop}%
\bibitem [{\citenamefont {Bustamante}\ \emph {et~al.}(2020)\citenamefont
  {Bustamante}, \citenamefont {Alexander}, \citenamefont {Maciuba},\ and\
  \citenamefont {Kaiser}}]{Bustamante2020}%
  \BibitemOpen
  \bibfield  {author} {\bibinfo {author} {\bibfnamefont {C.}~\bibnamefont
  {Bustamante}}, \bibinfo {author} {\bibfnamefont {L.}~\bibnamefont
  {Alexander}}, \bibinfo {author} {\bibfnamefont {K.}~\bibnamefont {Maciuba}},\
  and\ \bibinfo {author} {\bibfnamefont {C.~M.}\ \bibnamefont {Kaiser}},\
  }\bibfield  {title} {\bibinfo {title} {Single-molecule studies of protein
  folding with optical tweezers},\ }\href
  {https://doi.org/10.1146/annurev-biochem-013118-111442} {\bibfield  {journal}
  {\bibinfo  {journal} {Annual Review of Biochemistry}\ }\textbf {\bibinfo
  {volume} {89}},\ \bibinfo {pages} {443} (\bibinfo {year} {2020})},\ \bibinfo
  {note} {pMID: 32569525}\BibitemShut {NoStop}%
\bibitem [{\citenamefont {Gebhardt}\ \emph {et~al.}(2010)\citenamefont
  {Gebhardt}, \citenamefont {Bornschl{\"o}gl},\ and\ \citenamefont
  {Rief}}]{Gebhardt2010}%
  \BibitemOpen
  \bibfield  {author} {\bibinfo {author} {\bibfnamefont {J.~C.~M.}\
  \bibnamefont {Gebhardt}}, \bibinfo {author} {\bibfnamefont {T.}~\bibnamefont
  {Bornschl{\"o}gl}},\ and\ \bibinfo {author} {\bibfnamefont {M.}~\bibnamefont
  {Rief}},\ }\bibfield  {title} {\bibinfo {title} {Full distance-resolved
  folding energy landscape of one single protein molecule},\ }\href
  {https://doi.org/10.1073/pnas.0909854107} {\bibfield  {journal} {\bibinfo
  {journal} {Proceedings of the National Academy of Sciences}\ }\textbf
  {\bibinfo {volume} {107}},\ \bibinfo {pages} {2013} (\bibinfo {year}
  {2010})}\BibitemShut {NoStop}%
\bibitem [{\citenamefont {Neupane}\ \emph {et~al.}(2016)\citenamefont
  {Neupane}, \citenamefont {Manuel},\ and\ \citenamefont
  {Woodside}}]{Neupane2016}%
  \BibitemOpen
  \bibfield  {author} {\bibinfo {author} {\bibfnamefont {K.}~\bibnamefont
  {Neupane}}, \bibinfo {author} {\bibfnamefont {A.~P.}\ \bibnamefont
  {Manuel}},\ and\ \bibinfo {author} {\bibfnamefont {M.~T.}\ \bibnamefont
  {Woodside}},\ }\bibfield  {title} {\bibinfo {title} {Protein folding
  trajectories can be described quantitatively by one-dimensional diffusion
  over measured energy landscapes},\ }\href {https://doi.org/10.1038/nphys3677}
  {\bibfield  {journal} {\bibinfo  {journal} {Nature Physics}\ }\textbf
  {\bibinfo {volume} {12}},\ \bibinfo {pages} {700} (\bibinfo {year}
  {2016})}\BibitemShut {NoStop}%
\bibitem [{\citenamefont {Rebane}\ \emph {et~al.}(2016)\citenamefont {Rebane},
  \citenamefont {Ma},\ and\ \citenamefont {Zhang}}]{Rebane2016}%
  \BibitemOpen
  \bibfield  {author} {\bibinfo {author} {\bibfnamefont {A.~A.}\ \bibnamefont
  {Rebane}}, \bibinfo {author} {\bibfnamefont {L.}~\bibnamefont {Ma}},\ and\
  \bibinfo {author} {\bibfnamefont {Y.}~\bibnamefont {Zhang}},\ }\bibfield
  {title} {\bibinfo {title} {Structure-based derivation of protein folding
  intermediates and energies from optical tweezers},\ }\href
  {https://doi.org/https://doi.org/10.1016/j.bpj.2015.12.003} {\bibfield
  {journal} {\bibinfo  {journal} {Biophysical Journal}\ }\textbf {\bibinfo
  {volume} {110}},\ \bibinfo {pages} {441} (\bibinfo {year}
  {2016})}\BibitemShut {NoStop}%
\bibitem [{\citenamefont {Sharma}\ \emph {et~al.}(2007)\citenamefont {Sharma},
  \citenamefont {Perisic}, \citenamefont {Peng}, \citenamefont {Cao},
  \citenamefont {Lam}, \citenamefont {Lu},\ and\ \citenamefont
  {Li}}]{Sharma2007}%
  \BibitemOpen
  \bibfield  {author} {\bibinfo {author} {\bibfnamefont {D.}~\bibnamefont
  {Sharma}}, \bibinfo {author} {\bibfnamefont {O.}~\bibnamefont {Perisic}},
  \bibinfo {author} {\bibfnamefont {Q.}~\bibnamefont {Peng}}, \bibinfo {author}
  {\bibfnamefont {Y.}~\bibnamefont {Cao}}, \bibinfo {author} {\bibfnamefont
  {C.}~\bibnamefont {Lam}}, \bibinfo {author} {\bibfnamefont {H.}~\bibnamefont
  {Lu}},\ and\ \bibinfo {author} {\bibfnamefont {H.}~\bibnamefont {Li}},\
  }\bibfield  {title} {\bibinfo {title} {Single-molecule force spectroscopy
  reveals a mechanically stable protein fold and the rational tuning of its
  mechanical stability},\ }\href {https://doi.org/10.1073/pnas.0700351104}
  {\bibfield  {journal} {\bibinfo  {journal} {Proceedings of the National
  Academy of Sciences}\ }\textbf {\bibinfo {volume} {104}},\ \bibinfo {pages}
  {9278} (\bibinfo {year} {2007})}\BibitemShut {NoStop}%
\bibitem [{\citenamefont {Zheng}\ \emph {et~al.}(2014)\citenamefont {Zheng},
  \citenamefont {Wang},\ and\ \citenamefont {Li}}]{Zheng2014}%
  \BibitemOpen
  \bibfield  {author} {\bibinfo {author} {\bibfnamefont {P.}~\bibnamefont
  {Zheng}}, \bibinfo {author} {\bibfnamefont {Y.}~\bibnamefont {Wang}},\ and\
  \bibinfo {author} {\bibfnamefont {H.}~\bibnamefont {Li}},\ }\bibfield
  {title} {\bibinfo {title} {Reversible unfolding–refolding of rubredoxin: A
  single-molecule force spectroscopy study},\ }\href
  {https://doi.org/https://doi.org/10.1002/anie.201408105} {\bibfield
  {journal} {\bibinfo  {journal} {Angewandte Chemie International Edition}\
  }\textbf {\bibinfo {volume} {53}},\ \bibinfo {pages} {14060} (\bibinfo {year}
  {2014})}\BibitemShut {NoStop}%
\bibitem [{\citenamefont {Goldman}\ \emph {et~al.}(2015)\citenamefont
  {Goldman}, \citenamefont {Kaiser}, \citenamefont {Milin}, \citenamefont
  {Righini}, \citenamefont {Tinoco},\ and\ \citenamefont
  {Bustamante}}]{Goldman2015}%
  \BibitemOpen
  \bibfield  {author} {\bibinfo {author} {\bibfnamefont {D.~H.}\ \bibnamefont
  {Goldman}}, \bibinfo {author} {\bibfnamefont {C.~M.}\ \bibnamefont {Kaiser}},
  \bibinfo {author} {\bibfnamefont {A.}~\bibnamefont {Milin}}, \bibinfo
  {author} {\bibfnamefont {M.}~\bibnamefont {Righini}}, \bibinfo {author}
  {\bibfnamefont {I.}~\bibnamefont {Tinoco}},\ and\ \bibinfo {author}
  {\bibfnamefont {C.}~\bibnamefont {Bustamante}},\ }\bibfield  {title}
  {\bibinfo {title} {Mechanical force releases nascent
  chain{\textendash}mediated ribosome arrest in vitro and in vivo},\ }\href
  {https://doi.org/10.1126/science.1261909} {\bibfield  {journal} {\bibinfo
  {journal} {Science}\ }\textbf {\bibinfo {volume} {348}},\ \bibinfo {pages}
  {457} (\bibinfo {year} {2015})}\BibitemShut {NoStop}%
\bibitem [{\citenamefont {Cecconi}\ \emph {et~al.}(2005)\citenamefont
  {Cecconi}, \citenamefont {Shank}, \citenamefont {Bustamante},\ and\
  \citenamefont {Marqusee}}]{Cecconi2005}%
  \BibitemOpen
  \bibfield  {author} {\bibinfo {author} {\bibfnamefont {C.}~\bibnamefont
  {Cecconi}}, \bibinfo {author} {\bibfnamefont {E.~A.}\ \bibnamefont {Shank}},
  \bibinfo {author} {\bibfnamefont {C.}~\bibnamefont {Bustamante}},\ and\
  \bibinfo {author} {\bibfnamefont {S.}~\bibnamefont {Marqusee}},\ }\bibfield
  {title} {\bibinfo {title} {Direct observation of the three-state folding of a
  single protein molecule},\ }\href {https://doi.org/10.1126/science.1116702}
  {\bibfield  {journal} {\bibinfo  {journal} {Science}\ }\textbf {\bibinfo
  {volume} {309}},\ \bibinfo {pages} {2057} (\bibinfo {year}
  {2005})}\BibitemShut {NoStop}%
\bibitem [{\citenamefont {Stigler}\ \emph {et~al.}(2011)\citenamefont
  {Stigler}, \citenamefont {Ziegler}, \citenamefont {Gieseke}, \citenamefont
  {Gebhardt},\ and\ \citenamefont {Rief}}]{Stigler2011}%
  \BibitemOpen
  \bibfield  {author} {\bibinfo {author} {\bibfnamefont {J.}~\bibnamefont
  {Stigler}}, \bibinfo {author} {\bibfnamefont {F.}~\bibnamefont {Ziegler}},
  \bibinfo {author} {\bibfnamefont {A.}~\bibnamefont {Gieseke}}, \bibinfo
  {author} {\bibfnamefont {J.~C.~M.}\ \bibnamefont {Gebhardt}},\ and\ \bibinfo
  {author} {\bibfnamefont {M.}~\bibnamefont {Rief}},\ }\bibfield  {title}
  {\bibinfo {title} {The complex folding network of single calmodulin
  molecules},\ }\href {https://doi.org/10.1126/science.1207598} {\bibfield
  {journal} {\bibinfo  {journal} {Science}\ }\textbf {\bibinfo {volume}
  {334}},\ \bibinfo {pages} {512} (\bibinfo {year} {2011})}\BibitemShut
  {NoStop}%
\bibitem [{\citenamefont {Stockmar}\ \emph {et~al.}(2016)\citenamefont
  {Stockmar}, \citenamefont {Kobitski},\ and\ \citenamefont
  {Nienhaus}}]{Stockmar2016}%
  \BibitemOpen
  \bibfield  {author} {\bibinfo {author} {\bibfnamefont {F.}~\bibnamefont
  {Stockmar}}, \bibinfo {author} {\bibfnamefont {A.~Y.}\ \bibnamefont
  {Kobitski}},\ and\ \bibinfo {author} {\bibfnamefont {G.~U.}\ \bibnamefont
  {Nienhaus}},\ }\bibfield  {title} {\bibinfo {title} {Fast folding dynamics of
  an intermediate state in rnase h measured by single-molecule fret},\ }\href
  {https://doi.org/10.1021/acs.jpcb.5b09336} {\bibfield  {journal} {\bibinfo
  {journal} {The Journal of Physical Chemistry B}\ }\textbf {\bibinfo {volume}
  {120}},\ \bibinfo {pages} {641} (\bibinfo {year} {2016})},\ \bibinfo {note}
  {pMID: 26747376}\BibitemShut {NoStop}%
\bibitem [{\citenamefont {Yu}\ \emph {et~al.}(2017)\citenamefont {Yu},
  \citenamefont {Siewny}, \citenamefont {Edwards}, \citenamefont {Sanders},\
  and\ \citenamefont {Perkins}}]{Yu2017}%
  \BibitemOpen
  \bibfield  {author} {\bibinfo {author} {\bibfnamefont {H.}~\bibnamefont
  {Yu}}, \bibinfo {author} {\bibfnamefont {M.~G.~W.}\ \bibnamefont {Siewny}},
  \bibinfo {author} {\bibfnamefont {D.~T.}\ \bibnamefont {Edwards}}, \bibinfo
  {author} {\bibfnamefont {A.~W.}\ \bibnamefont {Sanders}},\ and\ \bibinfo
  {author} {\bibfnamefont {T.~T.}\ \bibnamefont {Perkins}},\ }\bibfield
  {title} {\bibinfo {title} {Hidden dynamics in the unfolding of individual
  bacteriorhodopsin proteins},\ }\href
  {https://doi.org/10.1126/science.aah7124} {\bibfield  {journal} {\bibinfo
  {journal} {Science}\ }\textbf {\bibinfo {volume} {355}},\ \bibinfo {pages}
  {945} (\bibinfo {year} {2017})}\BibitemShut {NoStop}%
\bibitem [{\citenamefont {Evans}\ and\ \citenamefont
  {Ritchie}(1997)}]{Evans1997}%
  \BibitemOpen
  \bibfield  {author} {\bibinfo {author} {\bibfnamefont {E.}~\bibnamefont
  {Evans}}\ and\ \bibinfo {author} {\bibfnamefont {K.}~\bibnamefont
  {Ritchie}},\ }\bibfield  {title} {\bibinfo {title} {Dynamic strength of
  molecular adhesion bonds},\ }\href@noop {} {\bibfield  {journal} {\bibinfo
  {journal} {Biophysical Journal}\ }\textbf {\bibinfo {volume} {72}},\ \bibinfo
  {pages} {1541} (\bibinfo {year} {1997})}\BibitemShut {NoStop}%
\bibitem [{\citenamefont {Bell}(1978)}]{Bell1978}%
  \BibitemOpen
  \bibfield  {author} {\bibinfo {author} {\bibfnamefont {G.~I.}\ \bibnamefont
  {Bell}},\ }\bibfield  {title} {\bibinfo {title} {Models for the specific
  adhesion of cells to cells},\ }\href@noop {} {\bibfield  {journal} {\bibinfo
  {journal} {Science}\ }\textbf {\bibinfo {volume} {200}},\ \bibinfo {pages}
  {618} (\bibinfo {year} {1978})}\BibitemShut {NoStop}%
\bibitem [{\citenamefont {Merkel}\ \emph {et~al.}(1999)\citenamefont {Merkel},
  \citenamefont {Nassoy}, \citenamefont {Leung}, \citenamefont {Ritchie},\ and\
  \citenamefont {Evans}}]{Merkel1999}%
  \BibitemOpen
  \bibfield  {author} {\bibinfo {author} {\bibfnamefont {R.}~\bibnamefont
  {Merkel}}, \bibinfo {author} {\bibfnamefont {P.}~\bibnamefont {Nassoy}},
  \bibinfo {author} {\bibfnamefont {A.}~\bibnamefont {Leung}}, \bibinfo
  {author} {\bibfnamefont {K.}~\bibnamefont {Ritchie}},\ and\ \bibinfo {author}
  {\bibfnamefont {E.}~\bibnamefont {Evans}},\ }\bibfield  {title} {\bibinfo
  {title} {Energy landscapes of receptor--ligand bonds explored with dynamic
  force spectroscopy},\ }\href@noop {} {\bibfield  {journal} {\bibinfo
  {journal} {Nature}\ }\textbf {\bibinfo {volume} {397}},\ \bibinfo {pages}
  {50} (\bibinfo {year} {1999})}\BibitemShut {NoStop}%
\bibitem [{\citenamefont {Evans}(2001)}]{Evans2001}%
  \BibitemOpen
  \bibfield  {author} {\bibinfo {author} {\bibfnamefont {E.}~\bibnamefont
  {Evans}},\ }\bibfield  {title} {\bibinfo {title} {Probing the relation
  between force—lifetime—and chemistry in single molecular bonds},\
  }\href@noop {} {\bibfield  {journal} {\bibinfo  {journal} {Annual Review of
  Biophysics and Biomolecular structure}\ }\textbf {\bibinfo {volume} {30}},\
  \bibinfo {pages} {105} (\bibinfo {year} {2001})}\BibitemShut {NoStop}%
\bibitem [{\citenamefont {Martin}\ \emph {et~al.}(1999)\citenamefont {Martin},
  \citenamefont {Richard}, \citenamefont {Salem}, \citenamefont {Hartley},\
  and\ \citenamefont {Mauguen}}]{Martin1999}%
  \BibitemOpen
  \bibfield  {author} {\bibinfo {author} {\bibfnamefont {C.}~\bibnamefont
  {Martin}}, \bibinfo {author} {\bibfnamefont {V.}~\bibnamefont {Richard}},
  \bibinfo {author} {\bibfnamefont {M.}~\bibnamefont {Salem}}, \bibinfo
  {author} {\bibfnamefont {R.}~\bibnamefont {Hartley}},\ and\ \bibinfo {author}
  {\bibfnamefont {Y.}~\bibnamefont {Mauguen}},\ }\bibfield  {title} {\bibinfo
  {title} {{Refinement and structural analysis of barnase at 1.5{\AA}
  resolution}},\ }\href {https://doi.org/10.1107/S0907444998010865} {\bibfield
  {journal} {\bibinfo  {journal} {Acta Crystallographica Section D}\ }\textbf
  {\bibinfo {volume} {55}},\ \bibinfo {pages} {386} (\bibinfo {year}
  {1999})}\BibitemShut {NoStop}%
\bibitem [{\citenamefont {Forns}\ \emph {et~al.}(2011)\citenamefont {Forns},
  \citenamefont {de Lorenzo}, \citenamefont {Manosas}, \citenamefont
  {Hayashi}, \citenamefont {Huguet},\ and\ \citenamefont {Ritort}}]{Forns2011}%
  \BibitemOpen
  \bibfield  {author} {\bibinfo {author} {\bibfnamefont {N.}~\bibnamefont
  {Forns}}, \bibinfo {author} {\bibfnamefont {S.}~\bibnamefont {de Lorenzo}},
  \bibinfo {author} {\bibfnamefont {M.}~\bibnamefont {Manosas}}, \bibinfo
  {author} {\bibfnamefont {K.}~\bibnamefont {Hayashi}}, \bibinfo {author}
  {\bibfnamefont {J.}~\bibnamefont {Huguet}},\ and\ \bibinfo {author}
  {\bibfnamefont {F.}~\bibnamefont {Ritort}},\ }\bibfield  {title} {\bibinfo
  {title} {Improving signal/noise resolution in single-molecule experiments
  using molecular constructs with short handles},\ }\href
  {https://doi.org/https://doi.org/10.1016/j.bpj.2011.01.071} {\bibfield
  {journal} {\bibinfo  {journal} {Biophysical Journal}\ }\textbf {\bibinfo
  {volume} {100}},\ \bibinfo {pages} {1765} (\bibinfo {year}
  {2011})}\BibitemShut {NoStop}%
\bibitem [{\citenamefont {Heidarsson}\ \emph {et~al.}(2014)\citenamefont
  {Heidarsson}, \citenamefont {Naqvi}, \citenamefont {Otazo}, \citenamefont
  {Mossa}, \citenamefont {Kragelund},\ and\ \citenamefont
  {Cecconi}}]{Heidarsson2014}%
  \BibitemOpen
  \bibfield  {author} {\bibinfo {author} {\bibfnamefont {P.~O.}\ \bibnamefont
  {Heidarsson}}, \bibinfo {author} {\bibfnamefont {M.~M.}\ \bibnamefont
  {Naqvi}}, \bibinfo {author} {\bibfnamefont {M.~R.}\ \bibnamefont {Otazo}},
  \bibinfo {author} {\bibfnamefont {A.}~\bibnamefont {Mossa}}, \bibinfo
  {author} {\bibfnamefont {B.~B.}\ \bibnamefont {Kragelund}},\ and\ \bibinfo
  {author} {\bibfnamefont {C.}~\bibnamefont {Cecconi}},\ }\bibfield  {title}
  {\bibinfo {title} {Direct single-molecule observation of calcium-dependent
  misfolding in human neuronal calcium sensor-1},\ }\href
  {https://doi.org/10.1073/pnas.1401065111} {\bibfield  {journal} {\bibinfo
  {journal} {Proceedings of the National Academy of Sciences}\ }\textbf
  {\bibinfo {volume} {111}},\ \bibinfo {pages} {13069} (\bibinfo {year}
  {2014})}\BibitemShut {NoStop}%
\bibitem [{\citenamefont {Jahn}\ \emph {et~al.}(2018)\citenamefont {Jahn},
  \citenamefont {Tych}, \citenamefont {Girstmair}, \citenamefont {Steinmaßl},
  \citenamefont {Hugel}, \citenamefont {Buchner},\ and\ \citenamefont
  {Rief}}]{Jahn2018}%
  \BibitemOpen
  \bibfield  {author} {\bibinfo {author} {\bibfnamefont {M.}~\bibnamefont
  {Jahn}}, \bibinfo {author} {\bibfnamefont {K.}~\bibnamefont {Tych}}, \bibinfo
  {author} {\bibfnamefont {H.}~\bibnamefont {Girstmair}}, \bibinfo {author}
  {\bibfnamefont {M.}~\bibnamefont {Steinmaßl}}, \bibinfo {author}
  {\bibfnamefont {T.}~\bibnamefont {Hugel}}, \bibinfo {author} {\bibfnamefont
  {J.}~\bibnamefont {Buchner}},\ and\ \bibinfo {author} {\bibfnamefont
  {M.}~\bibnamefont {Rief}},\ }\bibfield  {title} {\bibinfo {title} {Folding
  and domain interactions of three orthologs of hsp90 studied by
  single-molecule force spectroscopy},\ }\href
  {https://doi.org/https://doi.org/10.1016/j.str.2017.11.023} {\bibfield
  {journal} {\bibinfo  {journal} {Structure}\ }\textbf {\bibinfo {volume}
  {26}},\ \bibinfo {pages} {96} (\bibinfo {year} {2018})}\BibitemShut {NoStop}%
\bibitem [{\citenamefont {Mehlich}\ \emph {et~al.}(2020)\citenamefont
  {Mehlich}, \citenamefont {Fang}, \citenamefont {Pelz}, \citenamefont {Li},\
  and\ \citenamefont {Stigler}}]{Mehlich2020}%
  \BibitemOpen
  \bibfield  {author} {\bibinfo {author} {\bibfnamefont {A.}~\bibnamefont
  {Mehlich}}, \bibinfo {author} {\bibfnamefont {J.}~\bibnamefont {Fang}},
  \bibinfo {author} {\bibfnamefont {B.}~\bibnamefont {Pelz}}, \bibinfo {author}
  {\bibfnamefont {H.}~\bibnamefont {Li}},\ and\ \bibinfo {author}
  {\bibfnamefont {J.}~\bibnamefont {Stigler}},\ }\bibfield  {title} {\bibinfo
  {title} {Slow transition path times reveal a complex folding barrier in a
  designed protein},\ }\bibfield  {journal} {\bibinfo  {journal} {Frontiers in
  Chemistry}\ }\textbf {\bibinfo {volume} {8}},\ \href
  {https://doi.org/10.3389/fchem.2020.587824} {10.3389/fchem.2020.587824}
  (\bibinfo {year} {2020})\BibitemShut {NoStop}%
\bibitem [{\citenamefont {Shank}\ \emph {et~al.}(2010)\citenamefont {Shank},
  \citenamefont {Cecconi}, \citenamefont {Dill}, \citenamefont {Marqusee},\
  and\ \citenamefont {Bustamante}}]{Shank2010}%
  \BibitemOpen
  \bibfield  {author} {\bibinfo {author} {\bibfnamefont {E.~A.}\ \bibnamefont
  {Shank}}, \bibinfo {author} {\bibfnamefont {C.}~\bibnamefont {Cecconi}},
  \bibinfo {author} {\bibfnamefont {J.~W.}\ \bibnamefont {Dill}}, \bibinfo
  {author} {\bibfnamefont {S.}~\bibnamefont {Marqusee}},\ and\ \bibinfo
  {author} {\bibfnamefont {C.}~\bibnamefont {Bustamante}},\ }\bibfield  {title}
  {\bibinfo {title} {The folding cooperativity of a protein is controlled by
  its chain topology},\ }\href {https://doi.org/10.1038/nature09021} {\bibfield
   {journal} {\bibinfo  {journal} {Nature}\ }\textbf {\bibinfo {volume}
  {465}},\ \bibinfo {pages} {637 } (\bibinfo {year} {2010})}\BibitemShut
  {NoStop}%
\bibitem [{\citenamefont {Motlagh}\ \emph {et~al.}(2016)\citenamefont
  {Motlagh}, \citenamefont {Toptygin}, \citenamefont {Kaiser},\ and\
  \citenamefont {Hilser}}]{Motlagh2016}%
  \BibitemOpen
  \bibfield  {author} {\bibinfo {author} {\bibfnamefont {H.~N.}\ \bibnamefont
  {Motlagh}}, \bibinfo {author} {\bibfnamefont {D.}~\bibnamefont {Toptygin}},
  \bibinfo {author} {\bibfnamefont {C.~M.}\ \bibnamefont {Kaiser}},\ and\
  \bibinfo {author} {\bibfnamefont {V.~J.}\ \bibnamefont {Hilser}},\ }\bibfield
   {title} {\bibinfo {title} {Single-molecule chemo-mechanical spectroscopy
  provides structural identity of folding intermediates},\ }\href
  {https://doi.org/https://doi.org/10.1016/j.bpj.2015.12.042} {\bibfield
  {journal} {\bibinfo  {journal} {Biophysical Journal}\ }\textbf {\bibinfo
  {volume} {110}},\ \bibinfo {pages} {1280} (\bibinfo {year}
  {2016})}\BibitemShut {NoStop}%
\bibitem [{\citenamefont {Alemany}\ \emph {et~al.}(2016)\citenamefont
  {Alemany}, \citenamefont {Rey-Serra}, \citenamefont {Frutos}, \citenamefont
  {Cecconi},\ and\ \citenamefont {Ritort}}]{Alemany2016}%
  \BibitemOpen
  \bibfield  {author} {\bibinfo {author} {\bibfnamefont {A.}~\bibnamefont
  {Alemany}}, \bibinfo {author} {\bibfnamefont {B.}~\bibnamefont {Rey-Serra}},
  \bibinfo {author} {\bibfnamefont {S.}~\bibnamefont {Frutos}}, \bibinfo
  {author} {\bibfnamefont {C.}~\bibnamefont {Cecconi}},\ and\ \bibinfo {author}
  {\bibfnamefont {F.}~\bibnamefont {Ritort}},\ }\bibfield  {title} {\bibinfo
  {title} {Mechanical folding and unfolding of protein barnase at the
  single-molecule level},\ }\href
  {https://doi.org/https://doi.org/10.1016/j.bpj.2015.11.015} {\bibfield
  {journal} {\bibinfo  {journal} {Biophysical Journal}\ }\textbf {\bibinfo
  {volume} {110}},\ \bibinfo {pages} {63} (\bibinfo {year} {2016})}\BibitemShut
  {NoStop}%
\bibitem [{\citenamefont {Liu}\ \emph {et~al.}(2019)\citenamefont {Liu},
  \citenamefont {Chen},\ and\ \citenamefont {Kaiser}}]{Liu2019}%
  \BibitemOpen
  \bibfield  {author} {\bibinfo {author} {\bibfnamefont {K.}~\bibnamefont
  {Liu}}, \bibinfo {author} {\bibfnamefont {X.}~\bibnamefont {Chen}},\ and\
  \bibinfo {author} {\bibfnamefont {C.~M.}\ \bibnamefont {Kaiser}},\ }\bibfield
   {title} {\bibinfo {title} {Energetic dependencies dictate folding mechanism
  in a complex protein},\ }\href {https://doi.org/10.1073/pnas.1914366116}
  {\bibfield  {journal} {\bibinfo  {journal} {Proceedings of the National
  Academy of Sciences}\ }\textbf {\bibinfo {volume} {116}},\ \bibinfo {pages}
  {25641} (\bibinfo {year} {2019})}\BibitemShut {NoStop}%
\bibitem [{\citenamefont {Palassini}\ and\ \citenamefont
  {Ritort}(2011)}]{Pala2011}%
  \BibitemOpen
  \bibfield  {author} {\bibinfo {author} {\bibfnamefont {M.}~\bibnamefont
  {Palassini}}\ and\ \bibinfo {author} {\bibfnamefont {F.}~\bibnamefont
  {Ritort}},\ }\bibfield  {title} {\bibinfo {title} {Improving free-energy
  estimates from unidirectional work measurements: theory and experiment},\
  }\href@noop {} {\bibfield  {journal} {\bibinfo  {journal} {Phys. Rev. Lett.}\
  }\textbf {\bibinfo {volume} {107}},\ \bibinfo {pages} {060601} (\bibinfo
  {year} {2011})}\BibitemShut {NoStop}%
\bibitem [{\citenamefont {Rico-Pasto}\ \emph {et~al.}(2018)\citenamefont
  {Rico-Pasto}, \citenamefont {Pastor},\ and\ \citenamefont
  {Ritort}}]{Rico2018}%
  \BibitemOpen
  \bibfield  {author} {\bibinfo {author} {\bibfnamefont {M.}~\bibnamefont
  {Rico-Pasto}}, \bibinfo {author} {\bibfnamefont {I.}~\bibnamefont {Pastor}},\
  and\ \bibinfo {author} {\bibfnamefont {F.}~\bibnamefont {Ritort}},\
  }\bibfield  {title} {\bibinfo {title} {Force feedback effects on single
  molecule hopping and pulling experiments},\ }\href@noop {} {\bibfield
  {journal} {\bibinfo  {journal} {The Journal of Chemical Physics}\ }\textbf
  {\bibinfo {volume} {148}},\ \bibinfo {pages} {123327} (\bibinfo {year}
  {2018})}\BibitemShut {NoStop}%
\bibitem [{\citenamefont {Leffler}(1953)}]{Leffler1953}%
  \BibitemOpen
  \bibfield  {author} {\bibinfo {author} {\bibfnamefont {J.~E.}\ \bibnamefont
  {Leffler}},\ }\bibfield  {title} {\bibinfo {title} {Parameters for the
  description of transition states},\ }\href@noop {} {\bibfield  {journal}
  {\bibinfo  {journal} {Science}\ }\textbf {\bibinfo {volume} {117}},\ \bibinfo
  {pages} {340} (\bibinfo {year} {1953})}\BibitemShut {NoStop}%
\bibitem [{\citenamefont {Hammond}(1955)}]{Hammond1955}%
  \BibitemOpen
  \bibfield  {author} {\bibinfo {author} {\bibfnamefont {G.~S.}\ \bibnamefont
  {Hammond}},\ }\bibfield  {title} {\bibinfo {title} {A correlation of reaction
  rates},\ }\href@noop {} {\bibfield  {journal} {\bibinfo  {journal} {Journal
  of the American Chemical Society}\ }\textbf {\bibinfo {volume} {77}},\
  \bibinfo {pages} {334} (\bibinfo {year} {1955})}\BibitemShut {NoStop}%
\bibitem [{\citenamefont {van~der Sleen}\ and\ \citenamefont
  {Tych}(2021)}]{Tych2021}%
  \BibitemOpen
  \bibfield  {author} {\bibinfo {author} {\bibfnamefont {L.~M.}\ \bibnamefont
  {van~der Sleen}}\ and\ \bibinfo {author} {\bibfnamefont {K.~M.}\ \bibnamefont
  {Tych}},\ }\bibfield  {title} {\bibinfo {title} {Bioconjugation strategies
  for connecting proteins to dna-linkers for single-molecule force-based
  experiments},\ }\href@noop {} {\bibfield  {journal} {\bibinfo  {journal}
  {Nanomaterials}\ }\textbf {\bibinfo {volume} {11}},\ \bibinfo {pages} {2424}
  (\bibinfo {year} {2021})}\BibitemShut {NoStop}%
\bibitem [{\citenamefont {Huguet}\ \emph {et~al.}(2010)\citenamefont {Huguet},
  \citenamefont {Bizarro}, \citenamefont {Forns}, \citenamefont {Smith},
  \citenamefont {Bustamante},\ and\ \citenamefont {Ritort}}]{Huguet2010}%
  \BibitemOpen
  \bibfield  {author} {\bibinfo {author} {\bibfnamefont {J.~M.}\ \bibnamefont
  {Huguet}}, \bibinfo {author} {\bibfnamefont {C.~V.}\ \bibnamefont {Bizarro}},
  \bibinfo {author} {\bibfnamefont {N.}~\bibnamefont {Forns}}, \bibinfo
  {author} {\bibfnamefont {S.~B.}\ \bibnamefont {Smith}}, \bibinfo {author}
  {\bibfnamefont {C.}~\bibnamefont {Bustamante}},\ and\ \bibinfo {author}
  {\bibfnamefont {F.}~\bibnamefont {Ritort}},\ }\bibfield  {title} {\bibinfo
  {title} {Single-molecule derivation of salt dependent base-pair free energies
  in dna},\ }\href@noop {} {\bibfield  {journal} {\bibinfo  {journal} {PNAS}\
  }\textbf {\bibinfo {volume} {107}},\ \bibinfo {pages} {15431} (\bibinfo
  {year} {2010})}\BibitemShut {NoStop}%
\bibitem [{\citenamefont {de~Lorenzo}\ \emph {et~al.}(2015)\citenamefont
  {de~Lorenzo}, \citenamefont {Ribezzi-Crivellari}, \citenamefont
  {Arias-Gonzalez}, \citenamefont {Smith},\ and\ \citenamefont
  {Ritort}}]{DeLorenzo2015}%
  \BibitemOpen
  \bibfield  {author} {\bibinfo {author} {\bibfnamefont {S.}~\bibnamefont
  {de~Lorenzo}}, \bibinfo {author} {\bibfnamefont {M.}~\bibnamefont
  {Ribezzi-Crivellari}}, \bibinfo {author} {\bibfnamefont {J.~R.}\ \bibnamefont
  {Arias-Gonzalez}}, \bibinfo {author} {\bibfnamefont {S.~B.}\ \bibnamefont
  {Smith}},\ and\ \bibinfo {author} {\bibfnamefont {F.}~\bibnamefont
  {Ritort}},\ }\bibfield  {title} {\bibinfo {title} {A temperature-jump optical
  trap for single-molecule manipulation},\ }\href
  {https://doi.org/https://doi.org/10.1016/j.bpj.2015.05.017} {\bibfield
  {journal} {\bibinfo  {journal} {Biophysical Journal}\ }\textbf {\bibinfo
  {volume} {108}},\ \bibinfo {pages} {2854} (\bibinfo {year}
  {2015})}\BibitemShut {NoStop}%
\bibitem [{\citenamefont {Li}\ \emph {et~al.}(2006)\citenamefont {Li},
  \citenamefont {Collin}, \citenamefont {Smith}, \citenamefont {Bustamante},\
  and\ \citenamefont {Tinoco}}]{Tinoco2006}%
  \BibitemOpen
  \bibfield  {author} {\bibinfo {author} {\bibfnamefont {P.~T.~X.}\
  \bibnamefont {Li}}, \bibinfo {author} {\bibfnamefont {D.}~\bibnamefont
  {Collin}}, \bibinfo {author} {\bibfnamefont {S.~B.}\ \bibnamefont {Smith}},
  \bibinfo {author} {\bibfnamefont {C.}~\bibnamefont {Bustamante}},\ and\
  \bibinfo {author} {\bibfnamefont {I.~J.}\ \bibnamefont {Tinoco}},\ }\bibfield
   {title} {\bibinfo {title} {Probing the mechanical folding kinetics of tar
  rna by hopping, force-jump, and force-ramp methods},\ }\href@noop {}
  {\bibfield  {journal} {\bibinfo  {journal} {Biophysical Journal}\ }\textbf
  {\bibinfo {volume} {130}},\ \bibinfo {pages} {250} (\bibinfo {year}
  {2006})}\BibitemShut {NoStop}%
\bibitem [{\citenamefont {Woodside}\ \emph {et~al.}(2006)\citenamefont
  {Woodside}, \citenamefont {Behnke-Parks}, \citenamefont {Larizadeh},
  \citenamefont {Travers}, \citenamefont {Herschlag},\ and\ \citenamefont
  {Block}}]{Woodside2006}%
  \BibitemOpen
  \bibfield  {author} {\bibinfo {author} {\bibfnamefont {M.~T.}\ \bibnamefont
  {Woodside}}, \bibinfo {author} {\bibfnamefont {W.~M.}\ \bibnamefont
  {Behnke-Parks}}, \bibinfo {author} {\bibfnamefont {K.}~\bibnamefont
  {Larizadeh}}, \bibinfo {author} {\bibfnamefont {K.}~\bibnamefont {Travers}},
  \bibinfo {author} {\bibfnamefont {D.}~\bibnamefont {Herschlag}},\ and\
  \bibinfo {author} {\bibfnamefont {S.~M.}\ \bibnamefont {Block}},\ }\bibfield
  {title} {\bibinfo {title} {Nanomechanical measurements of the
  sequence-dependent folding landscapes of single nucleic acid hairpins},\
  }\href {https://doi.org/10.1073/pnas.0511048103} {\bibfield  {journal}
  {\bibinfo  {journal} {Proceedings of the National Academy of Sciences}\
  }\textbf {\bibinfo {volume} {103}},\ \bibinfo {pages} {6190} (\bibinfo {year}
  {2006})},\ \Eprint
  {https://arxiv.org/abs/https://www.pnas.org/content/103/16/6190.full.pdf}
  {https://www.pnas.org/content/103/16/6190.full.pdf} \BibitemShut {NoStop}%
\bibitem [{\citenamefont {Gao}\ \emph {et~al.}(2011)\citenamefont {Gao},
  \citenamefont {Sirinakis},\ and\ \citenamefont {Zhang}}]{Gao2011}%
  \BibitemOpen
  \bibfield  {author} {\bibinfo {author} {\bibfnamefont {Y.}~\bibnamefont
  {Gao}}, \bibinfo {author} {\bibfnamefont {G.}~\bibnamefont {Sirinakis}},\
  and\ \bibinfo {author} {\bibfnamefont {Y.}~\bibnamefont {Zhang}},\ }\bibfield
   {title} {\bibinfo {title} {Highly anisotropic stability and folding kinetics
  of a single coiled coil protein under mechanical tension},\ }\href@noop {}
  {\bibfield  {journal} {\bibinfo  {journal} {J. Am. Chem. Soc.}\ }\textbf
  {\bibinfo {volume} {133}},\ \bibinfo {pages} {12749} (\bibinfo {year}
  {2011})}\BibitemShut {NoStop}%
\bibitem [{\citenamefont {Gao}\ \emph {et~al.}(2012)\citenamefont {Gao},
  \citenamefont {Zorman}, \citenamefont {Gundersen}, \citenamefont {Xi},
  \citenamefont {Ma}, \citenamefont {Sirinakis}, \citenamefont {Rothman},\ and\
  \citenamefont {Zhang}}]{Gao2012}%
  \BibitemOpen
  \bibfield  {author} {\bibinfo {author} {\bibfnamefont {Y.}~\bibnamefont
  {Gao}}, \bibinfo {author} {\bibfnamefont {S.}~\bibnamefont {Zorman}},
  \bibinfo {author} {\bibfnamefont {G.}~\bibnamefont {Gundersen}}, \bibinfo
  {author} {\bibfnamefont {Z.}~\bibnamefont {Xi}}, \bibinfo {author}
  {\bibfnamefont {L.}~\bibnamefont {Ma}}, \bibinfo {author} {\bibfnamefont
  {G.}~\bibnamefont {Sirinakis}}, \bibinfo {author} {\bibfnamefont {J.~E.}\
  \bibnamefont {Rothman}},\ and\ \bibinfo {author} {\bibfnamefont
  {Y.}~\bibnamefont {Zhang}},\ }\bibfield  {title} {\bibinfo {title} {Single
  reconstituted neuronal snare complexes zipper in three distinct stages},\
  }\href@noop {} {\bibfield  {journal} {\bibinfo  {journal} {Science}\ }\textbf
  {\bibinfo {volume} {337}},\ \bibinfo {pages} {1340} (\bibinfo {year}
  {2012})}\BibitemShut {NoStop}%
\bibitem [{\citenamefont {Yu}\ \emph {et~al.}(2012)\citenamefont {Yu},
  \citenamefont {Guptaa}, \citenamefont {Liua}, \citenamefont {Neupanea},
  \citenamefont {Brigleyb}, \citenamefont {Sosovab},\ and\ \citenamefont
  {Woodside}}]{Woodside2012}%
  \BibitemOpen
  \bibfield  {author} {\bibinfo {author} {\bibfnamefont {H.}~\bibnamefont
  {Yu}}, \bibinfo {author} {\bibfnamefont {A.~N.}\ \bibnamefont {Guptaa}},
  \bibinfo {author} {\bibfnamefont {X.}~\bibnamefont {Liua}}, \bibinfo {author}
  {\bibfnamefont {K.}~\bibnamefont {Neupanea}}, \bibinfo {author}
  {\bibfnamefont {A.~M.}\ \bibnamefont {Brigleyb}}, \bibinfo {author}
  {\bibfnamefont {I.}~\bibnamefont {Sosovab}},\ and\ \bibinfo {author}
  {\bibfnamefont {M.~T.}\ \bibnamefont {Woodside}},\ }\bibfield  {title}
  {\bibinfo {title} {Energy landscape analysis of native folding of the prion
  protein yields the diffusion constant, transition path time, and rates},\
  }\href@noop {} {\bibfield  {journal} {\bibinfo  {journal} {PNAS}\ }\textbf
  {\bibinfo {volume} {109}},\ \bibinfo {pages} {14452} (\bibinfo {year}
  {2012})}\BibitemShut {NoStop}%
\bibitem [{\citenamefont {Hummer}\ and\ \citenamefont
  {Szabo}(2003)}]{Hummer2003}%
  \BibitemOpen
  \bibfield  {author} {\bibinfo {author} {\bibfnamefont {G.}~\bibnamefont
  {Hummer}}\ and\ \bibinfo {author} {\bibfnamefont {A.}~\bibnamefont {Szabo}},\
  }\bibfield  {title} {\bibinfo {title} {Kinetics from nonequilibrium
  single-molecule pulling experiments},\ }\href@noop {} {\bibfield  {journal}
  {\bibinfo  {journal} {Biophysical Journal}\ }\textbf {\bibinfo {volume}
  {85}},\ \bibinfo {pages} {5} (\bibinfo {year} {2003})}\BibitemShut {NoStop}%
\bibitem [{\citenamefont {Dudko}\ \emph {et~al.}(2006)\citenamefont {Dudko},
  \citenamefont {Hummer},\ and\ \citenamefont {Szabo}}]{Dudko2006}%
  \BibitemOpen
  \bibfield  {author} {\bibinfo {author} {\bibfnamefont {O.~K.}\ \bibnamefont
  {Dudko}}, \bibinfo {author} {\bibfnamefont {G.}~\bibnamefont {Hummer}},\ and\
  \bibinfo {author} {\bibfnamefont {A.}~\bibnamefont {Szabo}},\ }\bibfield
  {title} {\bibinfo {title} {Intrinsic rates and activation free energies from
  single-molecule pulling experiments},\ }\href
  {https://doi.org/10.1103/PhysRevLett.96.108101} {\bibfield  {journal}
  {\bibinfo  {journal} {Phys. Rev. Lett.}\ }\textbf {\bibinfo {volume} {96}},\
  \bibinfo {pages} {108101} (\bibinfo {year} {2006})}\BibitemShut {NoStop}%
\bibitem [{\citenamefont {Bustamante}\ \emph {et~al.}(1994)\citenamefont
  {Bustamante}, \citenamefont {Marko}, \citenamefont {Siggia},\ and\
  \citenamefont {Smith}}]{Siggia1994}%
  \BibitemOpen
  \bibfield  {author} {\bibinfo {author} {\bibfnamefont {C.}~\bibnamefont
  {Bustamante}}, \bibinfo {author} {\bibfnamefont {J.}~\bibnamefont {Marko}},
  \bibinfo {author} {\bibfnamefont {E.}~\bibnamefont {Siggia}},\ and\ \bibinfo
  {author} {\bibfnamefont {S.}~\bibnamefont {Smith}},\ }\bibfield  {title}
  {\bibinfo {title} {Entropic elasticity of lambda-phage dna},\ }\href
  {https://doi.org/10.1126/science.8079175} {\bibfield  {journal} {\bibinfo
  {journal} {Science}\ }\textbf {\bibinfo {volume} {265}},\ \bibinfo {pages}
  {1599} (\bibinfo {year} {1994})}\BibitemShut {NoStop}%
\bibitem [{\citenamefont {Viader-Godoy}\ \emph {et~al.}(2021)\citenamefont
  {Viader-Godoy}, \citenamefont {Manosas},\ and\ \citenamefont
  {Ritort}}]{ViaderGodoy2021}%
  \BibitemOpen
  \bibfield  {author} {\bibinfo {author} {\bibfnamefont {X.}~\bibnamefont
  {Viader-Godoy}}, \bibinfo {author} {\bibfnamefont {M.}~\bibnamefont
  {Manosas}},\ and\ \bibinfo {author} {\bibfnamefont {F.}~\bibnamefont
  {Ritort}},\ }\bibfield  {title} {\bibinfo {title} {Sugar-pucker force-induced
  transition in single-stranded dna},\ }\bibfield  {journal} {\bibinfo
  {journal} {International Journal of Molecular Sciences}\ }\textbf {\bibinfo
  {volume} {22}},\ \href {https://doi.org/10.3390/ijms22094745}
  {10.3390/ijms22094745} (\bibinfo {year} {2021})\BibitemShut {NoStop}%
\bibitem [{\citenamefont {Alemany}\ and\ \citenamefont
  {Ritort}(2014)}]{alemany2014}%
  \BibitemOpen
  \bibfield  {author} {\bibinfo {author} {\bibfnamefont {A.}~\bibnamefont
  {Alemany}}\ and\ \bibinfo {author} {\bibfnamefont {F.}~\bibnamefont
  {Ritort}},\ }\bibfield  {title} {\bibinfo {title} {Determination of the
  elastic properties of short ssdna molecules by mechanically folding and
  unfolding dna hairpins},\ }\href@noop {} {\bibfield  {journal} {\bibinfo
  {journal} {Biopolymers}\ }\textbf {\bibinfo {volume} {101}},\ \bibinfo
  {pages} {1193} (\bibinfo {year} {2014})}\BibitemShut {NoStop}%
\bibitem [{\citenamefont {Makarov}\ \emph {et~al.}(1993)\citenamefont
  {Makarov}, \citenamefont {Protasevich}, \citenamefont {Kuznetsova},
  \citenamefont {Fedorov}, \citenamefont {Korolev}, \citenamefont
  {Struminskaya}, \citenamefont {Bazhulina}, \citenamefont {Leshchinskaya},
  \citenamefont {Hartley}, \citenamefont {Kirpichnikov}, \citenamefont
  {Yakovlev},\ and\ \citenamefont {Esipova}}]{Makarov1993}%
  \BibitemOpen
  \bibfield  {author} {\bibinfo {author} {\bibfnamefont {A.~A.}\ \bibnamefont
  {Makarov}}, \bibinfo {author} {\bibfnamefont {I.~I.}\ \bibnamefont
  {Protasevich}}, \bibinfo {author} {\bibfnamefont {N.~V.}\ \bibnamefont
  {Kuznetsova}}, \bibinfo {author} {\bibfnamefont {B.~B.}\ \bibnamefont
  {Fedorov}}, \bibinfo {author} {\bibfnamefont {S.~V.}\ \bibnamefont
  {Korolev}}, \bibinfo {author} {\bibfnamefont {N.~K.}\ \bibnamefont
  {Struminskaya}}, \bibinfo {author} {\bibfnamefont {N.~P.}\ \bibnamefont
  {Bazhulina}}, \bibinfo {author} {\bibfnamefont {I.~B.}\ \bibnamefont
  {Leshchinskaya}}, \bibinfo {author} {\bibfnamefont {R.~W.}\ \bibnamefont
  {Hartley}}, \bibinfo {author} {\bibfnamefont {M.~P.}\ \bibnamefont
  {Kirpichnikov}}, \bibinfo {author} {\bibfnamefont {G.~I.}\ \bibnamefont
  {Yakovlev}},\ and\ \bibinfo {author} {\bibfnamefont {N.~G.}\ \bibnamefont
  {Esipova}},\ }\bibfield  {title} {\bibinfo {title} {Comparative study of
  thermostability and structure of close homologues - barnase and binase},\
  }\href {https://doi.org/10.1080/07391102.1993.10508695} {\bibfield  {journal}
  {\bibinfo  {journal} {Journal of Biomolecular Structure and Dynamics}\
  }\textbf {\bibinfo {volume} {10}},\ \bibinfo {pages} {1047} (\bibinfo {year}
  {1993})},\ \bibinfo {note} {pMID: 8357541}\BibitemShut {NoStop}%
\bibitem [{\citenamefont {Griko}\ \emph {et~al.}(1994)\citenamefont {Griko},
  \citenamefont {Makhatadze}, \citenamefont {Privalov},\ and\ \citenamefont
  {Hartley}}]{Griko1994}%
  \BibitemOpen
  \bibfield  {author} {\bibinfo {author} {\bibfnamefont {Y.~V.}\ \bibnamefont
  {Griko}}, \bibinfo {author} {\bibfnamefont {G.~I.}\ \bibnamefont
  {Makhatadze}}, \bibinfo {author} {\bibfnamefont {P.~L.}\ \bibnamefont
  {Privalov}},\ and\ \bibinfo {author} {\bibfnamefont {R.~W.}\ \bibnamefont
  {Hartley}},\ }\bibfield  {title} {\bibinfo {title} {Thermodynamics of barnase
  unfolding},\ }\href {https://doi.org/https://doi.org/10.1002/pro.5560030414}
  {\bibfield  {journal} {\bibinfo  {journal} {Protein Science}\ }\textbf
  {\bibinfo {volume} {3}},\ \bibinfo {pages} {669} (\bibinfo {year}
  {1994})}\BibitemShut {NoStop}%
\bibitem [{\citenamefont {Galano-Frutos}\ and\ \citenamefont
  {Sancho}(2019)}]{Galano2019}%
  \BibitemOpen
  \bibfield  {author} {\bibinfo {author} {\bibfnamefont {J.~J.}\ \bibnamefont
  {Galano-Frutos}}\ and\ \bibinfo {author} {\bibfnamefont {J.}~\bibnamefont
  {Sancho}},\ }\bibfield  {title} {\bibinfo {title} {Accurate calculation of
  barnase and snase folding energetics using short molecular dynamics
  simulations and an atomistic model of the unfolded ensemble: Evaluation of
  force fields and water models},\ }\href
  {https://doi.org/10.1021/acs.jcim.9b00430} {\bibfield  {journal} {\bibinfo
  {journal} {Journal of Chemical Information and Modeling}\ }\textbf {\bibinfo
  {volume} {59}},\ \bibinfo {pages} {4350} (\bibinfo {year} {2019})},\ \bibinfo
  {note} {pMID: 31513394}\BibitemShut {NoStop}%
\bibitem [{\citenamefont {Hyeon}\ and\ \citenamefont
  {Thirumalai}(2006)}]{Changbong2006}%
  \BibitemOpen
  \bibfield  {author} {\bibinfo {author} {\bibfnamefont {C.}~\bibnamefont
  {Hyeon}}\ and\ \bibinfo {author} {\bibfnamefont {D.}~\bibnamefont
  {Thirumalai}},\ }\bibfield  {title} {\bibinfo {title} {Forced-unfolding and
  force-quench refolding of rna hairpins},\ }\href@noop {} {\bibfield
  {journal} {\bibinfo  {journal} {Biophys. Journal}\ }\textbf {\bibinfo
  {volume} {90}},\ \bibinfo {pages} {3410–3427} (\bibinfo {year}
  {2006})}\BibitemShut {NoStop}%
\bibitem [{\citenamefont {Manosas}\ \emph {et~al.}(2006)\citenamefont
  {Manosas}, \citenamefont {Collin},\ and\ \citenamefont
  {Ritort}}]{Manosas2006}%
  \BibitemOpen
  \bibfield  {author} {\bibinfo {author} {\bibfnamefont {M.}~\bibnamefont
  {Manosas}}, \bibinfo {author} {\bibfnamefont {D.}~\bibnamefont {Collin}},\
  and\ \bibinfo {author} {\bibfnamefont {F.}~\bibnamefont {Ritort}},\
  }\bibfield  {title} {\bibinfo {title} {Force-dependent folding and unfolding
  kinetics in dna hairpins reveals transition-state displacements along a
  single pathway},\ }\href@noop {} {\bibfield  {journal} {\bibinfo  {journal}
  {Phys. Rev. Lett.}\ }\textbf {\bibinfo {volume} {96}},\ \bibinfo {pages}
  {218301 1} (\bibinfo {year} {2006})}\BibitemShut {NoStop}%
\bibitem [{\citenamefont {Alemany}\ and\ \citenamefont
  {Ritort}(2017)}]{Alemany2017}%
  \BibitemOpen
  \bibfield  {author} {\bibinfo {author} {\bibfnamefont {A.}~\bibnamefont
  {Alemany}}\ and\ \bibinfo {author} {\bibfnamefont {F.}~\bibnamefont
  {Ritort}},\ }\bibfield  {title} {\bibinfo {title} {Force-dependent folding
  and unfolding kinetics in dna hairpins reveals transition-state displacements
  along a single pathway},\ }\href@noop {} {\bibfield  {journal} {\bibinfo
  {journal} {J. Phys. Chem. Lett.}\ }\textbf {\bibinfo {volume} {8}},\ \bibinfo
  {pages} {895} (\bibinfo {year} {2017})}\BibitemShut {NoStop}%
\bibitem [{\citenamefont {Matouschek}\ and\ \citenamefont
  {Fersht}(1993)}]{Matouschek1993}%
  \BibitemOpen
  \bibfield  {author} {\bibinfo {author} {\bibfnamefont {A.}~\bibnamefont
  {Matouschek}}\ and\ \bibinfo {author} {\bibfnamefont {A.}~\bibnamefont
  {Fersht}},\ }\bibfield  {title} {\bibinfo {title} {Applications of physical
  organic chemistry to engineered mutants of barnase: Hammond postulate
  behaviour in the transition state of protein folding},\ }\href@noop {}
  {\bibfield  {journal} {\bibinfo  {journal} {PNAS}\ }\textbf {\bibinfo
  {volume} {90}},\ \bibinfo {pages} {7814} (\bibinfo {year}
  {1993})}\BibitemShut {NoStop}%
\bibitem [{\citenamefont {Waite}(1983)}]{Waite1983}%
  \BibitemOpen
  \bibfield  {author} {\bibinfo {author} {\bibfnamefont {J.}~\bibnamefont
  {Waite}},\ }\bibfield  {title} {\bibinfo {title} {11 - quinone-tanned
  scleroproteins},\ }in\ \href
  {https://doi.org/https://doi.org/10.1016/B978-0-12-751401-7.50018-1} {\emph
  {\bibinfo {booktitle} {Metabolic Biochemistry and Molecular Biomechanics}}},\
  \bibinfo {editor} {edited by\ \bibinfo {editor} {\bibfnamefont {P.~W.}\
  \bibnamefont {HOCHACHKA}}}\ (\bibinfo  {publisher} {Academic Press},\
  \bibinfo {year} {1983})\ pp.\ \bibinfo {pages} {467--504}\BibitemShut
  {NoStop}%
\bibitem [{\citenamefont {Lu}\ \emph {et~al.}(2009)\citenamefont {Lu},
  \citenamefont {Yeung}, \citenamefont {Sieracki},\ and\ \citenamefont
  {Marshall}}]{Lu2009}%
  \BibitemOpen
  \bibfield  {author} {\bibinfo {author} {\bibfnamefont {Y.}~\bibnamefont
  {Lu}}, \bibinfo {author} {\bibfnamefont {N.}~\bibnamefont {Yeung}}, \bibinfo
  {author} {\bibfnamefont {N.}~\bibnamefont {Sieracki}},\ and\ \bibinfo
  {author} {\bibfnamefont {N.~M.}\ \bibnamefont {Marshall}},\ }\bibfield
  {title} {\bibinfo {title} {Design of functional metalloproteins},\ }\href
  {https://doi.org/10.1038/nature08304} {\bibfield  {journal} {\bibinfo
  {journal} {Nature}\ }\textbf {\bibinfo {volume} {460}},\ \bibinfo {pages}
  {855} (\bibinfo {year} {2009})}\BibitemShut {NoStop}%
\bibitem [{\citenamefont {Hayashi}\ and\ \citenamefont
  {Wu}(1990)}]{Hayashi1990}%
  \BibitemOpen
  \bibfield  {author} {\bibinfo {author} {\bibfnamefont {S.}~\bibnamefont
  {Hayashi}}\ and\ \bibinfo {author} {\bibfnamefont {H.~C.}\ \bibnamefont
  {Wu}},\ }\bibfield  {title} {\bibinfo {title} {Lipoproteins in bacteria},\
  }\href {https://doi.org/10.1007/BF00763177} {\bibfield  {journal} {\bibinfo
  {journal} {Journal of Bioenergetics and Biomembranes}\ }\textbf {\bibinfo
  {volume} {22}},\ \bibinfo {pages} {451} (\bibinfo {year} {1990})}\BibitemShut
  {NoStop}%
\end{thebibliography}%

\end{document}